\documentclass{aa}  
\usepackage{graphicx}
\usepackage{natbib}
\usepackage{txfonts}
\begin{document}
   \title{A deep all-sky census of the Hyades}
\author{S.~R\"{o}ser \inst{1} \and
E.~Schilbach  \inst{1} \and
A.E.~Piskunov \inst{1,2,3}\and
 N.V.~Kharchenko \inst{1,3,4}\and
  R.-D.~Scholz \inst{3}}

   \offprints{S.~R\"{o}ser}

\institute{Astronomisches Rechen-Institut, Zentrum f\"ur Astronomie der Universit\"at Heidelberg, M\"{o}nchhofstra\ss{}e 12-14,
D--69120 Heidelberg, Germany\\
email: roeser@ari.uni-heidelberg.de, elena@ari.uni-heidelberg.de,
apiskunov@ari.uni-heidelberg.de, nkhar@ari.uni-heidelberg.de,
\and
Institute of Astronomy of the Russian Acad. Sci., 48 Pyatnitskaya
Str., 109017 Moscow, Russia\\
email: piskunov@inasan.rssi.ru
\and
Astrophysikalisches Institut Potsdam, An der Sternwarte 16, D--14482
Potsdam, Germany\\
email: apiskunov@aip.de,nkharchenko@aip.de,  rdscholz@aip.de 
\and
Main Astronomical Observatory, 27 Academica Zabolotnogo Str., 03680
Kiev, Ukraine\\
email: nkhar@mao.kiev.ua
}
   \date{Received March 24, 2011; accepted May 12, 2011}

 
  \abstract
  {} 
   {On the basis of the PPMXL catalogue we perform an all-sky census of the Hyades
   down to masses of about 0.2 $\rm{m}_\odot $ in a region up to 30~pc from the cluster centre.}
   {We use the proper motions from PPMXL in the convergent
point method to determine probable kinematic members. From 2MASS photometry
and CMC14 $r'$-band photometry, we
derive empirical
colour-absolute magnitude diagrams and, finally,
determine photometric membership for all kinematic candidates. }
   {This is the first deep ($\rm{r'} \leq$ 17) all-sky survey of the Hyades
allowing a full three-dimensional analysis of the cluster.   
The survey is complete down to at least $M_{K_s} $ = 7.3 or 0.25 $\rm{m}_\odot $.
We find 724 stellar systems co-moving with the bulk Hyades space
velocity, which represent a total mass of 435 $\rm{m}_\odot $. The tidal radius
is about 9 pc, and 275 $\rm{m_\odot}$ (364 systems) are gravitationally bound. This is the
cluster proper. Its mass density profile is perfectly fitted by a Plummer model with a central
density of 2.21 $\rm{m_{\odot} pc^{-3}}$ and a core radius of $\rm{r_{co}}$ = 3.10 pc,
while the
half-mass radius is $\rm{r_h}$ = 4.1 pc.
There are another 100 $\rm{m}_\odot $ in a volume between one and two tidal radii (halo),
and another 60 $\rm{m}_\odot $ up to a distance of 30 pc from the centre. Strong mass
segregation is inherent in the cluster.
The present-day luminosity and mass functions are noticeably different in various
parts of the cluster (core, corona, halo, and co-movers). They are strongly evolved
compared to presently favoured initial mass functions. The analysis of the velocity
dispersion of the cluster shows that about 20\% of its members must be binaries. As a by-product, we find
that
presently available theoretical isochrones are not able to adequately describe the near-infrared
colour-absolute magnitude relation for those cluster stars that are less massive than about
0.6 $\rm{m}_\odot $.}
   {}
      
  \maketitle
%
\section{Introduction}
Without doubt, the Hyades are one of the best-studied open clusters in our Galaxy.
Its proximity to the Sun has made it an interesting target for centuries, so it is
impossible here to give appropriate credit to all the astronomers who worked on the
Hyades. With no claim to completeness, we mention here the studies by
\citet{1952BAN....11..385V},
\citet{1969AJ.....74....2V},
\citet{1975A&A....43..423P},
\citet{1975AJ.....80..379H},
\citet{1988AJ.....96..198G}, and
\citet{1992MNRAS.257..257R}.
Recently, \citet{2008MNRAS.388..495H} have surveyed some 275 deg$^2$ of
the Hyades based on a combination of 2MASS and UKIDSS observations.

Loosely speaking, we distinguish between pencil beam studies and those relying on an
all-sky survey. All the above-mentioned studies are pencil beam studies since they
only cover a limited field-of-view, even if it is as large as 275 deg$^2$ in the case
of \citet{2008MNRAS.388..495H}. On the sphere, this only corresponds to a distance
of 7.3 pc from the Hyades centre. Pencil beam studies are based on dedicated observations and may
comprise a variety of astrophysical parameters. They can only indirectly reveal the full
three-dimensional structure of the cluster. 

An all-sky survey allows studying the cluster in its full three-dimensional
extent, but one usually 
relies on the few available data entries in the survey, e.g. proper motions or photometry in
a few bands. Of these studies we mention the work by \citet{1991A&A...243..386S} using
the old PPM and FK5 catalogues and the papers based on the Hipparcos observations by
\citet{1998A&A...331...81P} and \citet{2001A&A...367..111D}. These all-sky studies
were restricted to stars brighter than about $V \approx $  11 to 12 (Hipparcos)
without being complete to this magnitude. 

As far as membership of individual stars in the Hyades cluster is concerned, the
results of the studies above were collected in the Prosser \& Stauffer data base
(presently available from J. Stauffer, priv. comm.)
some fifteen years ago.
Prosser \& Stauffer's  data base was assumed to be complete down to 0.1 $\rm{m}_\odot $
\citep{2008A&A...481..661B}. Much effort has since been dedicated to the search
for very low-mass (below 0.1 $\rm{m}_\odot $) Hyades candidates. These attempts
have been partly unsuccessful \citep{1999AJ....118..997G}, or resulted in the
confirmation of membership of only one M8.5 dwarf  \citep{1999AJ....117..343R}.
\citet{2008MNRAS.388..495H} report the detection of 12 L-dwarfs in a 275 deg$^2$ field of view,
and \citet{2008A&A...481..661B} describe the detection of two brown dwarfs in an
area of 16 deg$^2$ around the Hyades centre.

Using PPMXL \citep{2010AJ....139.2440R} we can, for the first time, 
extend previous studies by going as deep in magnitude as the pencil
beam studies and still performing an all-sky survey.
We supply a complete sample of candidates
within 30 pc from the centre, determine individual distances to the stars,
and so resolve the full
three-dimensional spatial structure of the Hyades,
and its immediate neighbourhood, from the most massive stars down to almost
0.1 $\rm{m}_\odot $  
by an application of the convergent
method to the proper motions in PPMXL.

The paper has the following structure: in the next section we briefly introduce
the observational dataset from PPMXL we are using. 
In section \ref{method} we describe
the convergent point method.
Then, in section \ref{member} we define our criteria for membership 
determination and discuss the amount of possible contamination of the sample.
We compare the empirically derived colour-absolute magnitude diagram of the Hyades
with theoretical isochrones in section \ref{theiso}.
In section \ref{Spatial} we reveal the spatial structure of the Hyades and
derive the shape parameters of the cluster. Section \ref{MLF} is devoted
to the present-day luminosity and mass functions.
Then we discuss the internal velocity dispersion in section \ref{veldisp}, and, finally
we summarise our results in section \ref {summary}.
\section{Observations}~\label{obs}
As primary data of observations we use the PPMXL catalogue \citep{2010AJ....139.2440R}.
PPMXL contains positions and proper motions on the ICRS for some 900 million stars (from
the brightest stars down to about $V \approx 20$) and covers the complete sky. PPMXL also gives low-accuracy photometry
from USNO-B1.0 \citep{2003AJ....125..984M}, a subset of 410 million stars contains 2MASS 
\citep{2006AJ....131.1163S} photometry. The typical individual mean errors of the proper motions
range from better than 2 mas/y for the brightest stars with Tycho-2 \citep{2000A&A...355L..27H} observations
to more than 10 mas/y
for the faintest stars in the region south of -30$^\circ$  declination.
For the 2.5 million brightest stars Tycho-2 $B,V$ photometry is available.\\

For this work we cross-matched PPMXL with UCAC3 \citep{2010AJ....139.2184Z} and CMC14 \citep{2006yCat.1304....0C}.
We combined the positions and proper motions of PPMXL with the positions from CMC14 and 
from  UCAC3. The latter had to be reconstructed because the original positions of UCAC3 are not published.
With these data we performed a weighted least-squares adjustment to derive new, improved, mean positions and proper motions.
In the following we refer to this subset of PPMXL as the Carlsberg-UCAC (CU) subset. It turned out that the
most important feature in the CU is the accurate photometry in at least one optical band from CMC14,
the $r'$-magnitudes in the SDSS system. CMC14 observed the sky in the declination range from -30 deg
to about +52 deg, and it 
is 95\% complete down to $r'=16.8$ and 80\% to 17.0. Then completeness drops rapidly (when compared to
SDSS). Its limiting magnitude is $r'=17.8$ \citep{2006yCat.1304....0C}. The CU contains some 140 million stars,
90 million of which have $r'$-magnitudes from CMC14.

The astrometric information for the $\approx$ 120 000 Hipparcos stars is taken from the new reduction of
the Hipparcos data by \citet{2007ASSL..350.....V} instead of the data contained in PPMXL, the cross-matches
with 2MASS are kept as are the $B,V$ magnitudes from Tycho-2 of the Hipparcos stars.
\section{Convergent point method}~\label{method}
For a nearby open cluster like the Hyades the convergent point method is a suitable
tool to determine membership when only proper motions and (for a representative subset)
radial velocities are available.
The method is textbook knowledge \citep[see, e.g.][]{1968stki.book.....S},
and an excellent recent description can be found in \citet{2009A&A...497..209V}. Hence, we will
not repeat a description here.

In this paper, we did not apply the convergent point method from scratch, but used the six-dimensional
phase space parameters for the Hyades cluster centre from Table 7 in \citet{2009A&A...497..209V}. 
We use the galactic
rectangular coordinate system $X, Y, Z$ with origin in the Sun, and axes pointing to the
Galactic Centre ($X$), to the direction of galactic rotation ($Y$), and the North 
Galactic Pole ($Z$). In galactic coordinates, we
adopt [$\rm{x}_c,\,\rm{y}_c,\,\rm{z}_c$] = [-43.1, 0.7, -17.3] pc
for the position
and [$\rm{u}_c,\,\rm{v}_c,\,\rm{w}_c$] = [-41.1, -19.2, -1.4]
$\rm{km s}^{-1}$ for the mean motion of the cluster centre.
In other words, the stars most accurately measured, Hipparcos stars from the new reduction
by \citet{2007ASSL..350.....V}, determine the position and the motion of the Hyades cluster. Then the direction of the
convergent point is also given, see Eq. 7 in \citet{2009A&A...497..209V}.

Given the phase space parameters from \citet{2009A&A...497..209V}, we search for all stars
in the sky (in this case the PPMXL catalogue, resp. the CU subset) which have proper motions consistent with the
given space velocity {\bf v$_c$} = [$\rm{u}_c,\rm{v}_c,\rm{w}_c$] of the cluster centre. For each line-of-sight ($\alpha,\delta$)
we can then express
the vector of the space velocity by one component $\rm{v}_r$ parallel to the line-of-sight, the radial
velocity, and one perpendicular to it $\rm{v}_t$, the tangential velocity (in the tangential plane).
It is easy to see that, for a Hyades member, the component of {\bf v$_c$} parallel to the line-of-sight only depends on ($\alpha,\delta$),
and not on the distance of a star from the Sun. Hence, the same is true for the component of {\bf v$_c$} in the
tangential plane. The tangential velocity can be split into one component v$_\parallel$ in the direction
to the convergent point, and a component v$_\perp$ perpendicular to it. The expectation value of the
latter is zero. 

The PPMXL catalogue gives for each star the positions ($\alpha,\delta$) and the proper motions
($\mu_{\alpha},\mu_{\delta}$), i.e. four of the six components of the phase space coordinates, the
radial velocity $\rm{v}_r$ and the distance d from the Sun being the remaining two. A necessary condition
for a star to be a kinematic member of the Hyades cluster is:
\begin{equation}
                       | {\bf {v_c}} - {\bf {v}} | \leq \epsilon ,
\end{equation}		      
where {\bf v} = [u,v,w] is the space velocity of a candidate and $\epsilon $ is a suitable bound.
As long as the radial velocity $\rm{v}_r$ is not measured, we adopt as $\rm{v}_r$ the projection of 
the cluster motion {\bf v$_c$} to the line-of-sight.

At each position ($\alpha,\delta$) the proper motions	      
($\mu_{\alpha},\mu_{\delta}$) can be rotated into the direction of v$_\parallel$ and v$_\perp$. We call
these motion components  $\mu_\parallel$ and $\mu_\perp$. The remaining parameter, distance d or parallax $\varpi$, is
given by
\begin{equation}                 
           \varpi = f \times \frac{\mu_\parallel}{\rm{v}_\parallel} 
\end{equation}	   
where $ f  = 4.74\, \rm{km s}^{-1} \rm{kpc}^{-1}$ and $\varpi$ is the so-called secular parallax. As the method fixes
 $\rm{v}_r$ as explained above and $\varpi$ from Eq. 2, the necessary condition (Eq. 1) for kinematic membership
shrinks to 
            $ | \rm{v}_\perp  | \leq \epsilon $. 	        	   
Since the expectation value of $\rm{v}_\perp$ is zero, $< (\rm{v}_\perp^2) >^\frac{1}{2} $  is a measure
for the one-dimensional velocity dispersion of the cluster.

The convergent point method predicts the radial velocity $\rm{v}_r$ and the distance d from the Sun for a cluster candidate.
A final confirmation as a member can be obtained by directly measuring $\rm{v}_r$ and $\varpi$. Other than from
trigonometric parallaxes, the distance
d from the Sun can be confirmed also by determining
photometric parallaxes. The CU subset of PPMXL catalogue gives 2MASS $J,H,K_s$-photometry plus
$r'$-magnitudes. For the
2.5 million brightest stars in PPMXL $B$ and $V$ magnitudes from Tycho-2 are available. These photometric
data are used to confirm or reject photometrically the parallaxes from the convergent point method.
The radial velocities $\rm{v}_r$ can only be confirmed by direct measurements.
\section{Membership determination}~\label{member}
\subsection{Kinematic membership}~\label{kin}
The Hyades have a tidal radius of about 10 pc \citep[see, e.g.][]{1998A&A...331...81P} and an average 
velocity dispersion in one dimension of 0.23 $\rm{km s}^{-1}$ \citep{1988AJ.....96..198G}.
In this paper we do not concentrate only on the gravitationally bound cluster itself, it is
also interesting to investigate the situation in the immediate surroundings.
There we expect to reveal former Hyades members which already left the cluster because of dynamical evolution.

Therefore, we allow a bound $\epsilon = 4\,\rm{kms}^{-1}$ for the velocity component v$_\perp$. Additionally, we have
to take into account that v$_\parallel$ goes to zero when ($\alpha,\delta$) approaches the convergent point. This led us 
to set an upper bound $\eta $ 
for the tangent of the angle between the proper motion vector and the direction to the convergent 
point

$ \frac{\rm{v}_\perp}{\rm{v}_\parallel} = \frac{\mu_\perp}{\mu_\parallel} \leq \eta  $.

At the cluster centre $\eta = 1/6 $ corresponds to  $ | \rm{v}_\perp  | =  4 \rm{kms}^{-1} $, so we
set this condition to hold everywhere in the cluster.

We also specified a third bound by requiring that the distance $\rm{r}_c$ of a candidate
from the cluster centre should be less than 30 pc which is sufficiently large compared to the tidal radius.

These kinematic selection criteria are fulfilled by 15757 stars out of the 140 million contained in the CU subset.
They are subject to the photometric selection described in the next section.
At this stage we excluded all candidate white dwarfs from further consideration. White dwarfs in the Hyades are at the limit
of being observed in 2MASS, and if so, the photometric accuracy is low. That prevents us from using photometric
distances 
of white dwarfs from the CU to check their predicted secular parallaxes. We will investigate the white dwarfs associated
with the Hyades in a separate study.

\subsection{Photometric membership}~\label{cmd}
In the following we will check if the kinematically selected stars populate allowed loci
in the colour-absolute-magnitude diagrams. As the candidates occupy
a large magnitude range, there is no unique CMD where all the stars have sufficiently 
accurate photometry. Stars brighter than $V = 10$ have good quality in $B$ and $V$ mainly from
Tycho-2 \citep{2000A&A...355L..27H}, or ASCC-2.5 \citep{2001KFNT...17..409K}.
On the other hand, stars fainter than about $K_s$ = 5 have good quality photometry from
2MASS and CMC14.

Fortunately, the interstellar reddenning towards the Hyades is very low.
In a critical review on reddening determinations of the Hyades
\citet{2006AJ....132.2453T} concludes that $E(B-V) \leq 1.0$~mmag, hence reddenning has been neglected 
in all future considerations.

   \begin{figure}[h!]
   \centering
   \includegraphics[bb=31 31 540 670,angle=-90,width=8cm,clip]{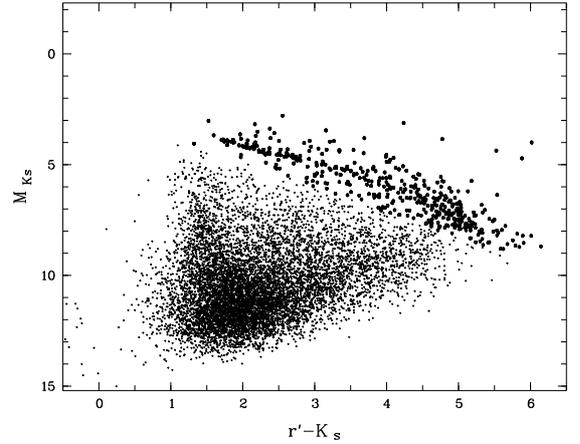}
      \caption{The $M_{K_s}$ vs. $r' - K_s$ diagram of the kinematic candidates (small dots).
      The vast majority of the stars have kinematically predicted parallaxes inconsistent
      with their photometric parallaxes. The candidates shown as thick dots are retained for further
      analysis. Stars brighter than  $M_{K_s} \approx 3.5 $ are not contained in CMC14.}
         \label{Fig1}
   \end{figure}
For the absolutely brightest stars we rely on the classical $M_V$ vs. $B - V$ diagram for
cross-checking the kinematically predicted parallax $\varpi$ with the photometric parallax. In fact, almost
all the bright Hyades members have been observed by Hipparcos and their membership is settled
\citep[see, e.g.][]{1998A&A...331...81P,2001A&A...367..111D,2009A&A...497..209V}.   

\begin{figure}[h!]
   \centering
   \includegraphics[bb=56 42 538 695,angle=-90,width=7cm,clip]{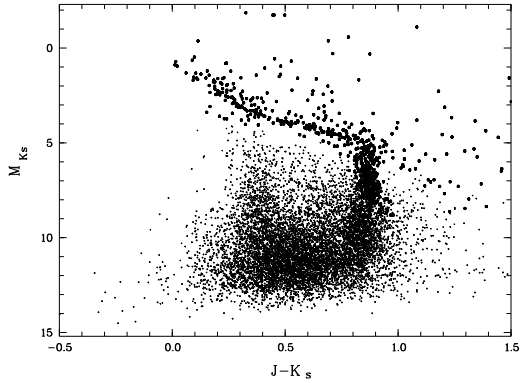}
\caption{The $M_{K_s}$ vs. $J - K_s$ diagram of the candidates. Stars which were marked
as thick dots in Fig. \ref{Fig1} are also marked as thick dots here.}
         \label{Fig2}
   \end{figure}

For stars with masses less than about 0.6\,$\rm{m}_\odot $ and the age and metallicity of the Hyades 
there are discrepancies between different theoretical isochrones,
especially in the NIR. We discuss this in more detail in Sect. \ref{theiso}.
As a consequence of this, we have no choice but to select
the Hyades candidates from an empirically derived colour-magnitude diagram obtained from the distribution
of the kinematically selected candidates. In the $M_{K_s}$ vs. $J - K_s$ diagram
the selection of low-mass stars ($ < 0.6\,\rm{m}_\odot $) is 
hampered by the fact that the colour $J - K_s$ is nearly constant.
In a first step, we therefore used the $M_{K_s}$ vs. $r' - K_s$ (Fig.\ref{Fig1}) and the $M_{K_s}$ vs. $J - K_s$
(Fig. \ref{Fig2}) diagrams to discard the large majority of the 15757 kinematic candidates whose
photometric parallaxes were inconsistent with the predicted secular ones.
   
In order to work out more clearly the empirical Hyades main sequence, we selected,
in Figs.\ref{Fig1} and \ref{Fig2}, only stars with
$ | \rm{v}_\perp  | \leq 1\,\rm{kms}^{-1}  $, i.e. the $\approx$9000 most probable kinematic candidates.
In Fig.\ref{Fig1}  we show the $M_{K_s}$ vs. $r' - K_s$ diagram of all those
9000 stars (small dots) that have $r'$ magnitudes, i.e. that are contained in CMC14.
The stars brighter than $M_{K_s} \approx 3.5 $ are missing. They are not contained
in CMC14, but their photometric membership is determined from Tycho-2 $B,V$ photometry.
The Hyades main sequence stands out prominently, and the candidates are provisionally marked 
as thicker dots. Stars below the Hyades sequence appear as sub-luminous stars and
must be much farther away than Hyades members. In other words, their 
parallax $\varpi$ predicted from the kinematic selection is inconsistent with the photometric
parallax, and hence these stars cannot belong to the Hyades. In the $M_{K_s}$ vs. $J - K_s$ diagram for
the same stars (Fig.\ref{Fig2}) we find a considerable density of stars at colours $J - K_s$ between
0.8 and 1. We marked the candidate Hyades from Fig. \ref{Fig1} as thick dots also in Fig. \ref{Fig2}.
It is clearly visible that NIR photometry alone is not sufficient to properly select faint stars with $M_{K_s} \ge 5$:
from Fig. \ref{Fig1} we find no Hyades candidates with $M_{K_s} > 9$  although Fig.\ref{Fig2} suggests plenty of them.
Instead, all the candidates with $M_{K_s} > 9$ have $r' - K_s$ colours which are to blue, and this means that their predicted
secular parallaxes are too large. We conclude that stars with $M_{K_s} > 9$ in the Hyades
are below the limiting magnitude of CMC14.

Even if we consider the finite width of the Hyades sequence caused by the presence
of binaries, we still find a number of
stars above the sequence in both CMDs (Fig.\ref{Fig1} and Fig. \ref{Fig2}).
In their majority they are field giants or very seldomly foreground dwarfs. 
The former are discarded via the two-colour, $J - H$ vs. $H - K_s$,
diagram.
In the $M_{K_s}$ vs. $J - K_s$  diagram we note a few white dwarfs near $J - K_s \approx 0$ and $11 \leq M_{K_s} \leq 14$, but 
investigating them needs a different approach.

For the final selection we determined the lower bound of an empirical cluster sequence stepwise over $r' - K_s$ and $J - K_s$,
and selected all stars between this lower bound and the upper limit given by binaries.
Out of the 15757  kinematically selected candidates, we could confirm the secular parallax via the
photometric parallax for 775 stars. 

\subsection{Final membership selection}~\label{finalmember}
The sample of 775 kinematic and photometric candidates was cross-matched with the probable Hyades members 
from the Prosser \& Stauffer database. Out of their 536 candidates, 407 were identified in our
sample. Further, we found 6 stars which belong to the Hyades with certainty but have been missing in our
list due to erroneous proper motions in PPMXL (checked via Vizier and Aladin). These stars were added to 
our sample to give a total of 781. The membership of the majority of the remaining stars
from the Prosser \& Stauffer  database has not been confirmed since they did not pass either the kinematic
or photometric criteria. A few stars from Prosser \& Stauffer are too faint to be
observed in CMC14, so we could not check their membership in a proper way.

\subsubsection{Contamination estimate}~\label{contest}
Although the kinematic and photometric criteria provide a
powerful tool for selecting cluster candidates and enable us to isolate 781 candidates out
of 140 million on the whole sky,
we cannot exclude a certain contamination of our
sample by field stars. The contamination can be estimated empirically by directly comparing the
predicted parallaxes and radial
velocities with the corresponding measurements as far as these are available.
Another approach is to compute the
probability of contamination from a kinematic model of the Galaxy, or, alternatively, by considering the 
observed velocity dispersion of field stars in the solar vicinity. In the following we apply these methods
to estimate probable contamination of our Hyades sample.

For about 300 stars from the candidate list we could find Hipparcos parallaxes and/or 
radial velocities in Vizier/Simbad. We call them ``control stars'' in the following. Whereas for the
majority of the control stars the predicted and measured parameters coincide well within the standard errors, 26
of them
have observed  parallaxes and/or radial velocities that differ significantly  from the computed ones. We analysed their
space coordinates predicted by the convergent point method and found that all of them are relatively far away
from the cluster centre. Indeed, the membership was not confirmed for all control stars with $|\rm{z} - \rm{z}_c| > 20$~pc
(11 stars).
The
remaining 15 ``wrong'' control stars are located at distances  $\rm{r}_c$ from the cluster centre larger than 9~pc.
From their distribution  
we estimate a 7.5\% contamination
at 9~pc $ <\rm{r}_c < 18$~pc, 30\% at 18~pc $ < \rm{r}_c < 30$~pc and 100\% whenever $|\rm{z} - \rm{z}_c| > 20$~pc.
Following this finding, we exclude all stars from the Hyades candidate sample which are located more than 20~pc away
from the cluster centre in z-direction. This leaves us with a sample of 724 candidates, and the
formal application
of the empirical rule found above predicts some 65 field stars among these 724 Hyades candidates.
   \begin{figure}[h!]
   \centering
   \includegraphics[bb=90 53 527 563,angle=-90,width=8cm,clip]{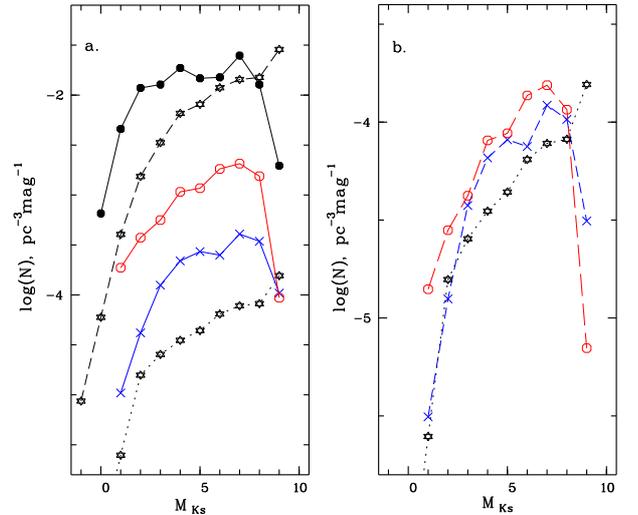}
\caption{The luminosity functions (LF) of the Hyades candidates and of
    field stars in the solar vicinity. {\bf a:} Solid lines and
symbols are the LFs of the Hyades at different distances from the cluster
centre $\rm{r}_c$,
       black filled circles for $\rm{r}_c < 9$~pc, red open circles for
9~pc $<\rm{r}_c < 18$~pc,
       blue crosses for 18~pc $< \rm{r}_c < 30$~pc. The LF of nearby stars
from the CNS4 (Jahrei{\ss}, H. , priv. comm.) is shown by the black dashed
line whereas the black
       dotted line is the fraction of field stars with
space velocities comparable
       to the velocity of the Hyades.
       {\bf b:} The red open circles and blue crosses show the adopted
contamination of the
       Hyades sample
       of 7.5\% at 9~pc $ <\rm{r}_c < 18$~pc and 30\% at 18~pc $< \rm{r}_c <
30$~pc, respectively.
       The black dotted line is the same as in {\bf a.} }
         \label{Fig3}
   \end{figure}

Since the control stars are relatively bright stars (see Fig.~\ref{Fig4}) that
usually have astrometric and photometric data of better accuracy,
we must consider the possibility that 
the contamination rate may be somewhat higher
for the faintest stars of our sample.
Therefore, we checked whether our 
empirical estimate from above coincides with the statistical predictions of
galactic models within the complete magnitude range covered by our sample. We used the kinematic model of the Galaxy by \citet{1996BaltA...5..337K}
to estimate the number of field stars within a circle of a radius 5~deg from the cluster centre
and with proper motions compatible with Hyades membership in this region of the sky. The model
predicts 5.06 such field stars in an area of 78.54~deg$^2$ in the direction of the Hyades. 
That suits perfectly the number of 4.95 stars we obtain if we use
the empirical rule of the contamination estimates from above.

Further, we computed the probability for field stars in the local solar neighbourhood to have, by chance, the 
same space velocities as the Hyades members. Jahrei{\ss} (priv. comm.) has recently determined
the K-band luminosity function of the stars in the solar neighbourhood from the CNS4,
the new 4th Catalog of Nearby Stars (in preparation) from which we infer the
volume density of the local sample. The CNS4 luminosity function
is shown in Fig.~\ref{Fig3}a together with the luminosity functions of the 724 Hyades candidates at different distances from the
cluster centre. The random contamination of our Hyades sample is now determined by the percentage of stars having the
same space velocity  $[\rm{u},\rm{v},\rm{w}]$ as the Hyades members with respect to the local standard of rest (LSR).
Corrected for the solar motion from \citet{2006A&A...445..545P},
the LSR velocity of the Hyades is [-31.8, -7.1, +6.1]. 
Assuming that the velocity
distribution of field stars follows a three-dimensional normal distribution with standard deviations given by
the velocity dispersions, we compute the number of field stars with Hyades motion in the range
$0 \le M_{K_s} \le 8$.
We used the data from
\citet{1998MNRAS.298..387D} who determined the velocity dispersions of field stars of different colours in the
solar neighbourhood from Hipparcos measurements.
Taking into account that only two velocity components are observed,
we find that the percentage of random field stars with Hyades motion is less 
than $0.6\times10^{-2}$ of the local density, and this is plotted
as the dotted line at the bottom of Fig.~\ref{Fig3}a. 

Since the spatial density of the Hyades candidates with $\rm{r}_c <$ 9~pc  is, on average, by two orders of magnitude 
higher than that expected from the local density of field stars, we conclude that contamination is negligible
for this volume. With increasing distance from the cluster centre, the density of the Hyades stars is decreasing,
and the probability of contamination becomes
higher. For comparison we show in Fig.\ref{Fig3}b the contribution of field stars
(the same as in Fig. \ref{Fig3}a) 
together with the contamination expected in outer regions, like what we
obtained empirically a few paragraphs above.
For 9~pc $<\rm{r}_c < 18$~pc, the contamination by field stars
is always less than the estimated 7.5\%, except for the faintest stars with $M_{K_s} > 8$ where our
catalogue becomes incomplete.
Also in the region 18~pc $<\rm{r}_c < 30$~pc, the contamination of 30\% from our empirical finding coincides well with the statistical
estimates in this magnitude range. 

To summarise: in the following, our study is based on a sample of 724 Hyades candidates. The contamination by field
stars is assumed to be dependent on the distance $\rm{r}_c$ from the cluster centre: a negligible contamination at
$\rm{r}_c < 9$~pc, 7.5\% at 9 pc $ <\rm{r}_c < 18$~pc, and 30\% at 18 pc$ < \rm{r}_c < 30$~pc. 
This is taken into account when we discuss the density distribution in the Hyades cluster, its luminosity and mass functions.

\subsubsection{The final sample}~\label{sample}      
The 724 Hyades candidates are listed in Table \ref{tab_hyades} organised as follows:
in the first column we give a running number. The next six columns are the right ascension and declination
for equinox and epoch J2000.0, the corresponding proper motions, and the mean errors of the proper motions. The following
12 columns (8 to 19) contain photometric data, i.e. magnitudes and their corresponding mean errors: Johnson $B$ and $V$, mainly from
ASCC-2.5, $J, H$ and $K_s$ from 2MASS, $r'$ from CMC14. All the following columns are derived quantities. 
Columns 20 and 21 are the secular parallax $\varpi$
and its mean error, column 22 gives the predicted distance from the cluster centre, column 23 is the predicted
radial velocity, columns 24 to 27 are $\rm{v}_\parallel$ and $\rm{v}_\perp $ with corresponding mean errors.
In the last column 28 we give the mass of the stars as derived in section \ref{Spatial} below.
This table is available from the CDS, Strasbourg, France.

\begin{table}[h!]
\caption{Description of the table containing the data of 724 Hyades candidates.
This table is available from the CDS, Strasbourg, France. }
\label{tab_hyades}
\setlength{\tabcolsep}{2pt}
\begin{tabular}{rlcl}
\hline
& Label&Units&Explanations\\
\hline
1  &name      &  ---    &  Running number \\
2  &RAdeg      &  deg   &  Right Ascension J2000.0, epoch 2000.0\\
3  &DEdeg       & deg   &  Declination J2000.0, epoch 2000.0\\
4  &pmRA        & mas/yr & Proper motion in RA$\cdot$cos(DE)\\
5  &pmDE        & mas/yr & Proper motion in DE\\
6  &$e_{pmRA}$  & mas/yr &  Mean error of pmRA$\cdot$cos(DE) \\
7  &$e_{pmDE}$  & mas/yr &  Mean error of pmDE\\
8  &$B$         & mag	& Johnson $B$ magnitude  from ASCC-2.5  \\
9  &$e_B$       & mag	&  Standard error of $B$ magnitude\\
10 &$V$         & mag	&  Johnson $V$ magnitude  from ASCC-2.5 \\
11 &$e_V$       & mag	&  Standard error of $V$ magnitude\\
12 &$J$         & mag	&  $J$ magnitude from 2MASS \\
13 &$e_J$       & mag	&  Standard error of $J$ magnitude\\
14 &$H$         & mag	&  $H$ magnitude from 2MASS \\
15 &$e_H$       & mag	&  Standard error of $H$ magnitude\\
16 &$K_s$         & mag	&  $K_s$ magnitude from 2MASS \\
17 &$e_{K_s}$       & mag	&  Standard error of $K_s$ magnitude\\
18 &$r'$         & mag	&  SDSS $r'$ magnitude  from CAMC14  \\
19 &$e_{r'}$       & mag	&  Standard error of $r'$ magnitude\\
20 &$\varpi$    & mas	&  Secular parallax\\
21 &$e_\varpi$    & mas	&  Mean error of Secular parallax\\
22 &$\rm{r_c}$    & pc	&  Distance from the cluster centre\\
23 &$\rm{v}_r$    & kms$^{-1}$ 	&  Predicted radial velocity\\
24 &$\rm{v}_\parallel$    & kms$^{-1}$ 	&  Tangential velocity in direction \\
   &    &  	& to the convergent point\\
25 &$e_{\rm{v}_\parallel}$    & kms$^{-1}$ 	&  Mean error of  $\rm{v}_\parallel$ \\
26 &$\rm{v}_\perp$    & kms$^{-1}$ 	&  Tangential velocity perpendicular to the direction \\
   &    &  	& to the convergent point\\ 
27 &$e_{\rm{v}_\perp}$    & kms$^{-1}$ 	&  Mean error of  $\rm{v}_\perp$ \\
28 &$\rm{m}$    & $\rm{m}_\odot$ 	&  mass of the star (from M/L relation,\\
   &    &  	& see Sect. \ref{Spatial})\\
\hline
\end{tabular}
\end{table}

\begin{figure}[h!]
   \centering
   \includegraphics[bb=60 60 540 553,angle=-90,width=8cm,clip]{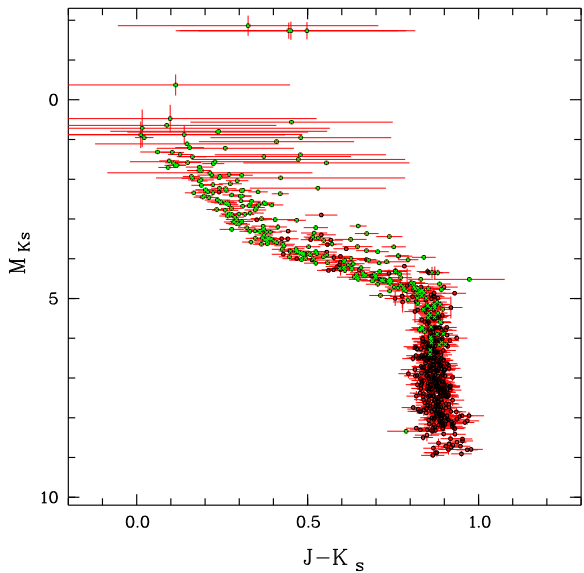}
      \caption{The $M_{K_s}$ vs. $J - K_s$ diagram of the 724 Hyades candidates. For $M_{K_s} \leq $ 2,
the $J - K_s$ colours from 2MASS are of low precision, because the stars are too bright in the
J-band. Stars shown as green dots are those for which we were able to confirm their membership by
comparing with measured trigonometric parallaxes and/or radial
velocities.}
         \label{Fig4}
   \end{figure}

The 
$M_{K_s}$ vs. $J - K_s$ colour-magnitude diagram of the final sample is shown in Fig. \ref{Fig4}.
Since the CU subset of PPMXL is 95\% complete down to $r'$ = 16.8, we find that our sample starts getting incomplete between
$M_{K_s}$ = 7 and $M_{K_s}$ = 8. A subset of stars in Fig. \ref{Fig4} is shown as green dots. For these stars we were
able to confirm their predicted secular parallaxes and/or radial velocities via measured trigonometric
parallaxes and/or radial velocities. 

\subsection{Comparison with theoretical isochrones}~\label{theiso}
The empirical selection of kinematic members was chosen as present-day theoretical isochrones
fail to describe the main sequence of the Hyades.   
We compared the empirical colour-magnitude diagram $M_{K_s}$ vs. $J - K_s$ with 
3 different theoretical isochrones.
In the literature, we found estimates of the metallicity of the Hyades, e.g. as [Fe/H]  = 0.14 $\pm$ 0.05 \citep{1998A&A...331...81P},
[Fe/H]  = 0.144 $\pm$ 0.013 \citep{2002HiA....12..680G} and [Fe/H]=0.13 $\pm$ 0.01 \citep{2003AJ....125.3185P}.
The age of the Hyades is given as  625 $\pm$ 50  Myr \citep{1998A&A...331...81P} or 648 $\pm$ 45, the latter based on cooling white dwarfs 
\citep{2009ApJ...696...12D}.  From models
taking into account stellar rotation of the turn-off stars  \citet{2001A&A...374..540L} estimate a range of 500 to 650 Myr
for the Hyades age.

\begin{figure}[h!]
   \centering
   \includegraphics[bb=89 41 578 692,angle=-90,width=7cm,clip]{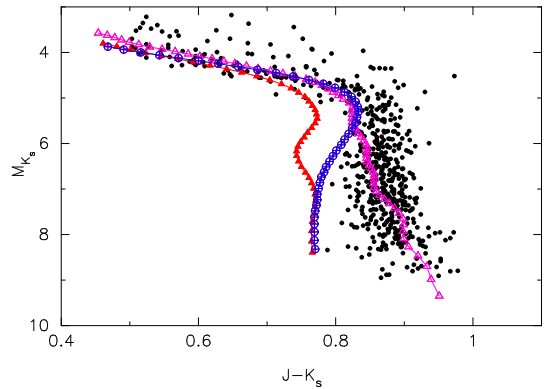}
     \caption{Comparison of the empirical $M_{K_s}$ vs. $J - K_s$ diagram of the
Hyades candidates (masses less than about 1 $\rm{m}_\odot $) with different theoretical isochrones.
The PADOVA isochrones are shown by filled red triangles (curve most to the left), the DARTMOUTH
isochrones blue open  circles, and the BCAH isochrones by the magenta open triangles (curve most to the right).} 

         \label{isocs}
   \end{figure}

In Fig. \ref{isocs}  we compare the empirical colour-magnitude diagram with available
theoretical isochrones for ages and metallicities as close to the Hyades values as possible. The isochrones are
taken from \citet{2008A&A...482..883M} (PADOVA, 650 Myr, [Fe/H]  = 0.14,
filled red triangles),  \citet{2008ApJS..178...89D}
(DARTMOUTH, 600 Myr, [Fe/H]  = 0.21, open blue circles ) and \citet{1998A&A...337..403B}
(BCAH, 650 Myr, [Fe/H]  = 0.0, open magenta triangles). In the latter case we used BCAH98\_models.1. The BCAH models are insensitive
to the L\_mix parameter below 0.6 $\rm{m}_\odot $ \citep{1998A&A...337..403B}. When isochrones were not given
in 2MASS magnitude bands, we converted them using the formulae given in the 2MASS webpage 
{\tt www.ipac.caltech.edu/2mass/}.

We find that the DARTMOUTH isochrones describe the $M_{K_s}$ vs. $J - K_s$
of the Hyades perfectly in the regime brighter than $M_{K_s} = 5.0 $, but fail to describe the situation for fainter stars.
This is, Hyades stars with masses less than 0.65 $\rm{m}_\odot $ 
have colours redder in $J - K_s$ than predicted by DARTMOUTH for stars with corresponding age and
 metallicity. The less massive the stars are, the greater the discrepancy.
The PADOVA isochrones  also coincide well with observations for $J - K_s \leq 0.7$ (0.7 $\rm{m}_\odot $).
For stars with masses less than 0.7 $\rm{m}_\odot $ the theoretical $J - K_s$ is about 0.1 mag too small.

Finally, a comparison has been made with the BCAH isoschrones.
They cover a mass range from 0.06 to 1.1 $\rm{m}_\odot $. 
For masses larger than 0.8 $\rm{m}_\odot $ ($J - K_s = 0.6$), BCAH gives $M_{K_s}$-magnitudes which
are slightly too bright compared to
PADOVA and DARTMOUTH. For  $4.5 \leq M_{K_s} \leq  6.5$ the theoretical
$J - K_s$-colours are slightly too blue compared to observations. But for fainter stars the 
isochrones 
satisfactory describe the observations.

We conclude that all the isochrones have problems with masses around 0.6
$\rm{m}_\odot $, which is at the transition from pre-MS to MS at the age of the Hyades
\citep[see, e.g][]{2000A&A...358..593S}

It is a mass-$M_V$ relation for masses above 1 $\rm{m}_\odot $,
and a mass-$M_{K_s}$ relation for masses lower than 1 $\rm{m}_\odot $.

\section{Spatial structure}~\label{Spatial} 
The analysis of the spatial structure of the cluster does not only require the three-dimensional distribution
of the stars as particles, but also their individual masses. v

For the 724 candidates we determined individual masses using the following
mass-to-luminosity relations:
for systems with $M_V \leq 5.38$ corresponding to m $\geq \rm{m}_\odot $ we take the Mass-$M_V$ relationship for the Hyades
derived by \citet{2004ApJ...600..946P}.  For $M_V > 5.38$ masses are determined via the Mass-$M_{K_s}$ relationship
from the DARTMOUTH isochrones \citep{2008ApJS..178...89D}. At the transition, we find an insignificant discontinuity
in the mass determination of less than
0.005 $ \rm{m}_\odot $. Finally for $M_{K_s} > 5.5$, masses are determined via the BCAH isochrones \citep{1998A&A...337..403B}.
The transition at $M_{K_s} = 5.5 $ creates an insignificant step in the estimated masses
of 0.01 $ \rm{m}_\odot $. In the range $4.0 \leq M_{K_s} \leq 6.0 $ masses determined by either of the isochrones
(DARTMOUTH or BCAH) do not differ by more than 0.03 $ \rm{m}_\odot $. 
In our application of
these mass-to-luminosity relations we ignore the possible binary nature of the stars in our sample, since
we have information on binarity only for a minor portion of our sample.  

\begin{figure}[h!]
   \centering
   \includegraphics[bb=102 55 538 705,angle=-90,width=7cm,clip]{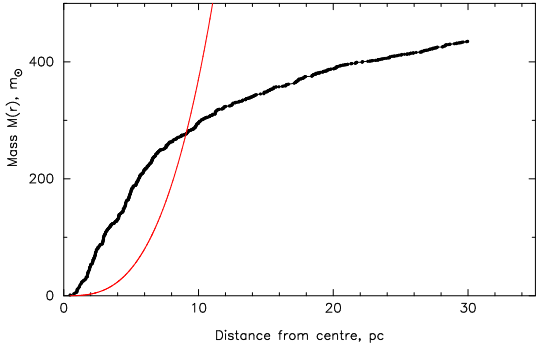}
      \caption{The cumulated mass as a function of the distance from the cluster centre $\rm{r_c}$.
      The thick black curve is the observed cumulated mass function as obtained from adding
      up all the individual masses within a distance $\rm{r_c}$. The (thinner) red curve is the relation
      from Eq. \ref{tidal} between the tidal radius and the tidal mass in the
      gravitational field of the Galaxy near the position of the Sun.}
         \label{mofr}
   \end{figure}

Using the individual masses of stellar systems, we derive the cumulative mass function M($\rm{r_c}$) depending on
the distance $\rm{r_c}$ from the cluster centre. This function is shown in Fig. \ref{mofr}.
It is corrected for contamination as described in Sec.~\ref{contest}. In the galactic disk the sphere of influence of a
gravitational body is given by
\begin{equation}                 
    \rm{x_L} = \left(\frac{G M_c}{4A(A-B)}\right)^\frac{1}{3} = \left(\frac{G M_c}{4\Omega_0^2 - \kappa^2}\right)^\frac{1}{3}
  \label{tidal}       
\end{equation}
where $\rm{x_L}$ is the distance of the Lagrangian points from the
centre, $\rm{M_c}$ is the total mass inside a distance
$\rm{x_L}$ from the centre, 
G = $4.3 \times 10^{-3} \rm{pc/m_\odot (km/s)^2}$ is the gravitational constant, A and B are 
Oort's constants, $\Omega_0$ the angular velocity and $\kappa$ the epicyclic frequency at
the position of the Sun. Here we use
 A (14.5 $\rm{kms}^{-1}\rm{kpc}^{-1}$) and
B (-13.0 $\rm{kms}^{-1} \rm{kpc}^{-1}$) from \citet{2006A&A...445..545P}.
The distance $\rm{x_L}$ is often referred to as the tidal radius $\rm{r_t}$ of a cluster and we
use this definition in the following also. The tidal radius $\rm{r_t}$ separates,
in general, stars gravitationally bound to a cluster from those that are unbound.
However, this definition is not to be understood that {\em each} star inside $\rm{r_t}$
is bound, and all stars outside are unbound. Individual stars can, every now and then,
change from a bound to an unbound state and vice versa
\citep[see][for the fraction of potential escapers with Jacobi energy above the critical value in the cluster]
{2009MNRAS.392..969J}.
It is inherent in the
convergent point method, that we find, in the first place, stars that are co-moving
with the adopted space motion of the centre of the Hyades cluster. So, we cannot separate,
from the beginning, the gravitationally bound Hyades cluster and its surroundings.

If we would add up the masses of all our Hyades co-moving stars within 30 pc from the
centre we would get a tidal radius $\rm{r_t}$ of 10.5 pc from Eq. \ref{tidal}.
Now, it is clear that stars outside 10.5 pc are not gravitationally bound, and cannot
self-consistently contribute to the determination of $\rm{r_t}$ itself. In general,
the tidal radius is the distance
$\rm{r_t}$ where $\rm{M(\rm{r_t}})$ calculated via Eq. \ref{tidal} equals the integrated
mass of all stars up to the distance $\rm{r_t}$ from the centre.
The function shown as the thin red curve in Fig. \ref{mofr} represents the relation from Eq.
\ref{tidal}, and the black curve gives the counted integrated mass of our stars as
a function of the distance from the centre. From Fig. \ref{mofr} we find that
the tidal radius comes out to be 9.0 pc and the
corresponding tidal mass is 276 $\rm{m}_{\odot}$. Uncertainty in the determination
of $\rm{r_t}$ may come from the fact that we use system masses in the case
of unresolved binaries, that we neglect the contribution from white dwarfs, and that our sample is getting incomplete at the
lowest masses. The additional mass in white dwarfs, found so far in the Hyades,
is very low. It has been estimated to be 6.4 $\rm{m}_\odot $ by \citet{1998AJ....115.1536V}.
We show in Sect. \ref{PDMF} that incompleteness at the low-mass end of the mass function
has only negligible influence on the tidal radius. We would miss less than 8\%
of the total mass in the unrealistic assumption that the logarithmic mass function were constant from 0.25 $\rm{m}_\odot $
down to 0.01 $\rm{m}_\odot$. 
The contribution due to unresolved binaries to the cluster mass is more substantial.
Based on their radial velocity study, \citet{1988AJ.....96..198G} argue that about
25\% of the stars they measured are binaries. Assuming the
average mass of the secondary of 2/3 of the primary's mass, \citet{1988AJ.....96..198G} estimate an
increase in the total mass
by 17\%. Applying this correction would give us an increase in the
tidal radius of only 5\%, from 9.0 pc to 9.45 pc. Even a 50\% increase in the total
mass would only enlarge the tidal radius to 10.4 pc.

Another source of uncertainty for the determination of the tidal radius
are the adopted values of Oort's constants to characterise the
tidal field of the Galaxy. Using different values for the parameters of galactic rotation, we could determine a tidal
radius between 8.0 pc (using $\Omega_0$ = 30.6, and $\kappa$ = 39.0, both in $\rm{kms}^{-1} \rm{kpc}^{-1}$)
 and 9.3 pc (using A = -B = 13.75 $\rm{kms}^{-1} \rm{kpc}^{-1}$) . The \citet{2006A&A...445..545P} values for A and B give a
tidal radius in between these two approaches. So,
for the rest of the paper,
we use a tidal radius of $\rm{r_t}$ = 9 pc. In models of star clusters,
especially in $N$-body calculations, the Lagrangian radius for 50\% of the cluster
mass, plays an important
role and is called 
half-mass radius $\rm{r_h}$. Estimated from a tidal mass of 276 $\rm{m}_{\odot}$, the half-mass radius is
$\rm{r_h}$ = 4.1 pc at the present state of evolution of the Hyades.

 \begin{figure}[h!]
   \centering
   \includegraphics[bb=41 33 578 725,angle=-90,width=10.5cm,clip]{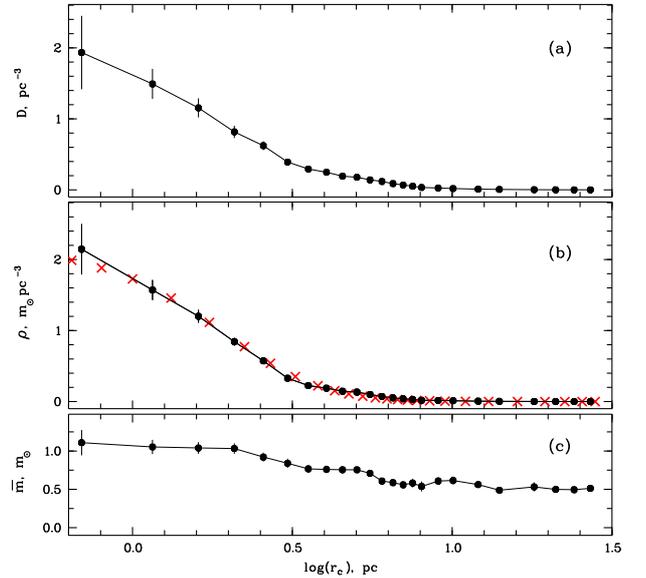}
      \caption{The dependence of number density, mass density, and average mass per star
      as a function of the logarithm of the distance from the centre  $\rm{r_c}$.
      From top to bottom:
     a) The number density D of stars in the cluster obtained with a binning of 2~ pc in steps of 1pc.
     Outside 10 pc the bin size is increased to get a better signal-to-noise ratio.
     The error bars
      are Poisson errors. b) The observed mass density $\rho$ (black dots) using the same binning
      and a density distribution (red crosses) from a Plummer model (Eq. \ref{pluden}
      with M$_t$ = 276  $\rm{m}_{\odot}$ and $\rm{r_{co}}$ = 3.10 pc).
      c) the average mass per star obtained as $\rho$/D. Its decrease 
      from the centre outbound characterises the mass segregation in the Hyades, at present time.}
         \label{plummer}
   \end{figure}

In Fig. \ref{plummer} we show the number density D, the mass density $\rho$ , and the
average mass per star as a function of the distance from the centre $\rm{r_{c}}$.
The bottom part of Fig. \ref{plummer} reveals the mass segregation in the Hyades;
the average mass per star decreases from 1.4 $\rm{m}_{\odot}$ in the centre to about
0.5 $\rm{m}_{\odot}$ at the tidal radius $\rm{r_t}$. We discuss this in more detail in Sect.
\ref{PDMF} together with the description of the mass function.
The mass density of the Hyades, shown in the middle part of Fig. \ref{plummer}, is fitted
to a Plummer model \citep{1915MNRAS..76..107P}, where the mass density follows the equation
\begin{equation}
\rho(\rm{r_{c}}) = \rm{\frac{3M_t}{4\pi{r_{co}}^3}\frac{1}{[1+(r_c/{r_{co}})^2]^{5/2}}},
\label{pluden}
\end{equation}
where $\rm{r_{co}}$ is the so-called core radius of a cluster. Using the tidal mass  of M$_t$ = 276  $\rm{m}_{\odot}$
the best fit to the observed density distribution (black dots) in Fig. \ref{plummer} is
obtained with a core radius of $\rm{r_{co}}$ = 3.10 pc.
The core radius is easily seen in the two upper panels of Fig. \ref{plummer} as the distance
where the slope in the density and mass distributions significantly changes.  
The
corresponding Plummer model (crosses in Fig. \ref{plummer}) shows excellent agreement
with the observations. The model has a central mass density of 2.21 $\rm{m_{\odot} pc^{-3}}$
which coincides well with our innermost point of 2.14 $\rm{m_{\odot} pc^{-3}}$. The ratio
of the half-mass radius $\rm{r_h}$ to $\rm{r_{co}}$ in the Hyades is 1.32 which also
is in remarkable coincidence with the theoretical ratio of 1.3048 
for a Plummer model \citep[see][]{1975ApJ...200..339S}.

\citet{1988AJ.....96..198G} already found that the cluster conforms well to a Plummer
model. Taking into account the incompleteness of their sample, they argue for a total
mass of 390 $\rm{m}_{\odot}$. They use the projected density distribution to determine a core
radius $\rm{r_{co}}$ = 3.15 pc (0.07 radians) and
a central density of 2.97 $\rm{m_{\odot} pc^{-3}}$, the latter being about 30\% higher than ours.
As the core radii $\rm{r_{co}}$ in Gunn's model and ours practically coincide, the
difference in central density comes solely via the tidal mass of the cluster.
So, either \citet{1988AJ.....96..198G} have overestimated the tidal mass of 390 $\rm{m}_{\odot}$,
or our stellar masses are underestimated essentially due to binarity. Using Gunn's estimate on binarity (25\%)
we could account for an increase of 17\% in total mass to get 323 $\rm{m}_{\odot}$. A mass of 390 $\rm{m}_{\odot}$
would require an increase of 37\% in mass for all our stellar systems (more than 50\% binaries),
which, however, could also be possible.
\citet{1998A&A...331...81P} also fit the observed density distribution to a Plummer model
and find a smaller core radius of 2.9 pc and a central density of 1.8 $\rm{m_{\odot} pc^{-3}}$.
This is not surprising as the Hipparcos sample is incomplete already at 0.6 to 0.7 $\rm{m}_{\odot}$.

In the following we call the region inside $\rm{r_{co}}$ = 3.1 pc the core, the region between
$\rm{r_{co}}$ and $\rm{r_t}$ the corona.
In a volume between 1 and 2 $\rm{r_t}$, called halo, we find another
100 $\rm{m}_{\odot}$ in objects co-moving with the Hyades, but probably not gravitationally bound.
Outside the halo up to the distance of 30 pc from the cluster centre there is 
another 60 $\rm{m}_{\odot}$ in co-moving stars.
   
\subsection{The shape of the cluster}~\label{nbody}
\begin{figure*}[t!]
   \centering
   \includegraphics[bb=45 405 558 646,angle=0,width=17cm,clip]{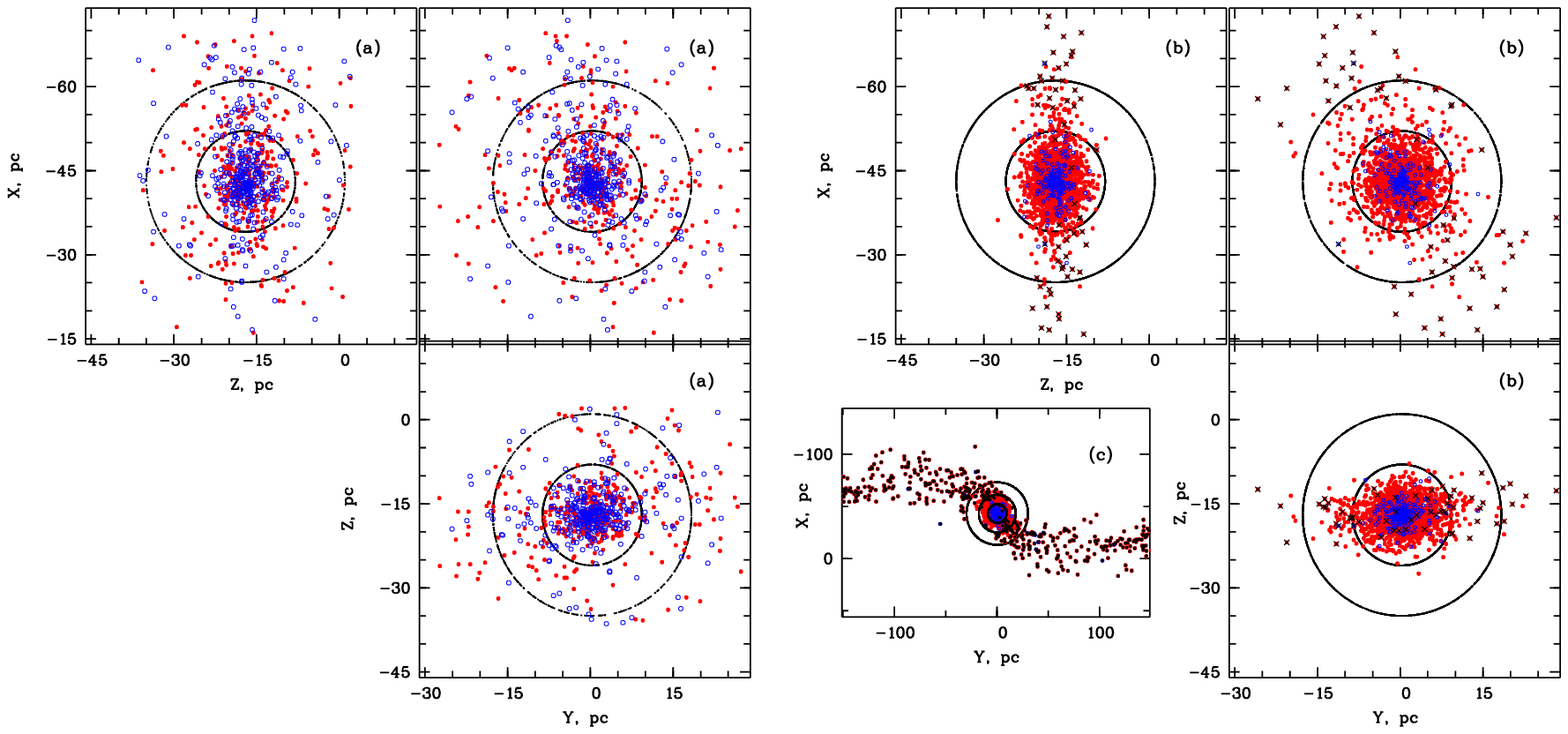}
      \caption{Observed (a) and modelled (b), (c) spatial distribution in the Hyades. 
   Blue points are for stars with 
   masses $\rm{m > 0.5m_{\odot}}$, 
   red points for $\rm{m \le 0.5m_{\odot}}$. In the modelled distribution, the dynamically 
   unbound particles are marked by black crosses (b) or black points  (c).
   Two large black circles in (a) and (b) indicate the
   distances $\rm{r}_c$ from the cluster centre of 9~pc and 18~pc, whereas in (c) the
   circles correspond to $\rm{r}_c=9$~pc, $\rm{r}_c=18$~pc,  and $\rm{r}_c=30$~pc. The last one restricts the area
considered in this study.
   }
         \label{xyz}
   \end{figure*}

Since the convergent point method provides an individual distance from the Sun for each
Hyades star, we can study the three-dimensional distribution of cluster members in more
detail. The distribution of the 724 Hyades candidates in galactic rectangular coordinates 
is shown in Fig.~\ref{xyz}a.

For comparison, we show in Fig.~\ref{xyz}b a model distribution of stars
predicted by the $N$-body simulation \citep{2009A&A...495..807K} of an open cluster
originally spheroidal, 
with an initial mass of 1000~$\rm{m}_{\odot}$ and the Salpeter IMF down to 0.1 $\rm{m}_{\odot}$.
The model cluster is moving on a circular orbit in 
the external tidal field  of the Milky Way. It is 650~Myr old and located
close to the observed location of the  Hyades \citep[for more details of the model parameters, see]
[]{2009A&A...495..807K}. It is appropriate to note that the model has not been specifically tailored to fit
the Hyades, but it is a general model for the evolution of a prototype open cluster in the disk at 8.5~kpc.
The model does not take into account that the Hyades could undergo some irregular
events (e.g., encounters with molecular clouds) during their life.
As far as the observations are concerned, we note that
our sample can be contaminated beyond $\rm{r}_c = 9$~pc.
However, the model and observations
show quite a good agreement in general.
The main feature, an elongated shape with the minor axis directed
towards the Galactic Poles, is clearly seen in both distributions. 

In the following we analyse and compare the shapes of the observed Hyades cluster and the
model cluster. The principal axes of  the cluster ellipsoid
are given as the eigenvectors and eigenvalues of the matrix formed by the 2nd order momenta  $M_{xx}, \cdots M_{zz}$
of the space coordinates of the cluster stars
\begin{center}

\[ \mathcal{M} = \left( \begin{array}{ccc}
M_{xx}  & M_{xy} & M_{xz} \\
M_{xy} & M_{yy}  & M_{yz} \\
M_{xz} & M_{yz} & M_{zz} 
\end{array} \right) .\]

\end{center}
The coordinates $x,y,z$ are counted relative to the
cluster centre in the galactic coordinate system as defined in Sect. \ref{method}.
Let $A,\,B,\,C$ be the eigenvalues of $\mathcal{M}$ with $A$ being the highest, $C$ the lowest eigenvalue.
The corresponding eigenvectors are denoted by  \overrightarrow{a}, \overrightarrow{b} and \overrightarrow{c}.

The results are given in Table \ref{ellip}. In the first column we indicate the data set used for the calculations
({\bf O}bservations or {\bf M}odel, within {\bf 30}~pc, {\bf 18}~pc, or {\bf 9}~pc). Columns 2 and 3
give the axis ratios. Columns 4, 5, and 6 describe the orientation of the ellipsoids in the $XYZ$-system
where $\psi$ is the angle between the $X$-axis and the projection of the semi-major axis \overrightarrow{a} 
on the $XZ$-plane, $\varphi$ is the angle between the $X$-axis and the projection of \overrightarrow{a} 
on the $XY$-plane, and $\theta$ is the angle between the $Y$-axis and the projection of \overrightarrow{b} 
on the $YZ$-plane. The numbers in brackets below each parameter give a measure of confidence.
These are the lowest and highest values obtained for a given parameter from four subsets of data in 
the corresponding range of $\rm{r}_c$. The first two subsets were extracted from the sample arranged
with increasing distance of stars from the cluster centre by sorting into two groups of even and odd 
entries. These
two subsets contain, therefore, half of stars of the original sample and are supplementing each other.
The other two subsets were extracted in the same way but from the sample arranged with increasing 
distance of stars from the Sun.

\begin{table}
\caption{Shape parameters of the observed Hyades {\bf{O}}, and the model cluster {\bf{M}}. See a detailed description in the text.}             
\label{table:1}      
\centering                          
\begin{tabular}{c c c c c c }        
\hline\hline                 
       &  $B/A$     &  $C/A$        & $\psi$, deg &$\varphi$, deg &$\theta$, deg\\    
\hline                        
{\bf O/30}   & 0.81       & 0.58       & 3        &  33       &  12  \\
       &[0.78; 0.84]&[0.55; 0.60]&[-1; 6]&[22; 42]&[4; 18]  \\
{\bf M/30}   & 0.71       & 0.44       & 1     &  30    &  4  \\
       &[0.70; 0.72]&[0.43; 0.45]&[-1; 2]&[25; 36]&[2;  7] \\
\hline
{\bf O/18}   & 0.86       & 0.64       & 2     &  27    &  19 \\
       &[0.81; 0.89]&[0.55; 0.70]&[-3; 5]&[12; 36]&[12; 31] \\     
{\bf M/18}   & 0.86       & 0.57       & 0     &  23    &  5      \\
       &[0.86; 0.87]&[0.55; 0.59]&[-1; 2]&[18; 28]&[3; 7] \\
\hline
{\bf O/9 }   & 0.89       &0.77        & 0     &  8      &  23  \\
       &[0.74; 0.93]&[0.75; 0.83]&[-8; 9]&[-18; 15]&[8; 42] \\
{\bf M/9 }   & 0.94       &0.78        & 0     &  22     &  7 \\
       &[0.91; 0.97]&[0.76; 0.79]&[-6; 8]&[3; 35]  &[0; 14] \\
\hline                                   
\end{tabular}
\label{ellip}
\end{table} 

Although the model particles are distributed more regularly than the real stars, both
distributions show similar parameters and tendencies. If the area within $\rm{r}_c = 30$~pc
is considered, the semi-major axis $A$ of the observed as well as of the modelled ellipsoids 
is about a factor of two larger than the semi-minor axis $C$ ($C/A = 0.58$ for O/30, and
$C/A = 0.44$ for M/30). The axis \overrightarrow{a} is located in the $XY$-plane ($\psi \approx 0^{\circ}$) 
but inclined to the $X$-axis by an angle $\varphi$ of about $30^{\circ}$, whereas the second 
axis \overrightarrow{b} seems to form
a small angle $\theta$ with respect to the  $XY$-plane. With decreasing $\rm{r}_c$ the distribution of stars  
becomes more spherical, and the orientation angles lose their meaning. 

An interesting feature in Fig. \ref{xyz}c, the tails parallel to the $Y$-axis, is predicted by the model 
\citep[see also Fig.~3 in]
[]{2009A&A...495..807K}. These tails spread out to $\approx 700$~pc from the
cluster centre and they are sparsely populated by former cluster members, mainly low-mass stars 
with space velocities pretty different to that of the cluster itself. Therefore, even if the 
tails would not be destroyed by events not taken into account by the model, they are
difficult to segregate from the field by use of kinematic criteria alone. From this point of view,
all-sky surveys of chemical compositions of stars would be more promising to detect former members
of an individual cluster.
 
\section{The mass and luminosity functions}~\label{MLF}
In this paper we derive the masses of our Hyades candidates (stellar systems) from 
a mass-luminosity relation (MLR). Hence the luminosity function and the mass function
are not independent of each other, and cannot be discussed
independently. From the point of view of an observer, the luminosities are the primary
observable quantities and the masses are derived thereof. However, from theory, the primary astrophysical
parameter is the initial mass function (IMF) which evolves into the present-day mass function (PDMF).
In consequence, we shall epitomise the observed luminosity function in the next subsection,
and discuss the PDMF in more detail.

In section \ref{Spatial} we described the mass-luminosity relation we adopted to convert
absolute magnitudes into masses. It is a mass-$M_V$ relation for masses above 1 $\rm{m}_\odot $,
and a mass-$M_{K_s}$ relation for masses lower than 1 $\rm{m}_\odot $.
Our derived luminosity and mass functions in this section were corrected for contamination as described in Sect.
\ref{contest} (7.5\% for 9~pc $ <\rm{r}_c < 18$~pc and 30\% at 18~pc $ < \rm{r}_c < 30$~pc).
In section \ref{cmd} we found that our sample is getting incomplete between
$M_{K_s}$ = 7.3 and $M_{K_s}$ = 8. Translated into mass, incompleteness in the mass functions
sets on somewhere between 0.25 and 0.17  $\rm{m}_\odot $. The onset of incompleteness is shown
by the dashed vertical lines in Figs. \ref{lfsiggi} and \ref{mf}.

\subsection{The luminosity function}~\label{PDLF}
The K-band luminosity function (KLF) 
is defined as the number of stars per magnitude bin and volume element
\begin{equation} 
 \Phi(M_{K_s}) = \frac{dN(M_{K_s})}{dM_{K_s}}.
\end{equation}
The KLF is mostly used to characterise the stellar population in
young or even embedded clusters. On the other hand, Jahrei{\ss} (priv. comm.) has recently determined
the KLF of the stars in the solar neighbourhood from the CNS4. We
compare the KLF from the Hyades both with that of young clusters and with the older local
population. 

In different regimes of magnitude, the KLF is usually characterised by a power law in the form
\begin{equation}
\frac{dN(M_{K_s})}{dM_{K_s}} \propto 10^{\alpha M_{K_s} }
\end{equation}
with $\alpha$ as the slope of the power law. For young open clusters \citep[see, e.g.][as an overview]
{2008AJ....135.2095D,2008MNRAS.384.1675J}
the slopes $\alpha$ of the KLF lie between  0.3 and 0.4 in ranges of $M_{K_s} $ between
0 and 3, whereas the
slope of the KLF of CNS4 (see Fig. \ref{Fig3}) is steeper, 
about 0.5, in the corresponding magnitude range.

\begin{figure}[h!]
   \centering
   \includegraphics[bb=69 40 513 322,angle=-90,width=6cm,clip]{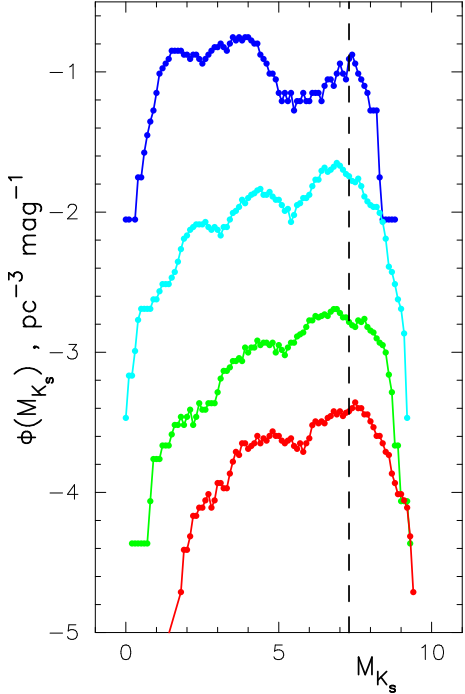}
      \caption{The luminosity function of the Hyades in different regions.
      This luminosity function is constructed as a moving average with a rectangular
      kernel of 1.5 magnitudes in steps of 0.1 mag. The upper (blue) curve
      refers to the core region inside 3.1 pc from the centre, the second (cyan) curve to the corona
      (3.1 pc to 9 pc), the third (green) to the halo (9 pc to 18 pc),
      and the fourth (red) to the co-movers outside 18 pc. The dashed
   vertical line at $M_{K_s}$ = 7.3 marks the onset of incompleteness of our survey.      
      The units are stars (systems) per $\rm{pc^{-3}}$ and 1 mag.}
         \label{lfsiggi}
   \end{figure}

To be independent of binning we construct the luminosity function in steps of 0.1 mag
with a binsize (kernel) of 1.5 mag. We also distinguish different regions of the cluster
as defined in Sec. \ref{Spatial},
the core region up to $\rm{r_{c}}$ = 3.1 pc from the centre, the corona between core and tidal radius of
9 pc, the halo in the 1 to 2 tidal radii shell, and the co-movers outside 18 pc up to
30 pc from the cluster centre. We show the KLF for the different regions of the Hyades in 
Fig. \ref{lfsiggi}.

Looking at the KLF of the different regions one feature immediately 
catches the eye: the luminosity function in the core of the Hyades is strikingly different from
the other regions and also from the field star LF in Fig. \ref{Fig3}.
We observe a strong excess of luminous stars with $M_{K_s} \leq 5$. This is
a manifestation of mass segregation already seen in Fig. \ref{plummer}. 

The maxima of the PDLFs in the different regions are reached at about $M_{K_s} $ = 7, except for the additional
maxima in the core at $M_{K_s} $ = 4 and $M_{K_s} $ = 1.5, caused by mass segregation.
The local KLF from Fig. \ref{Fig3} is still increasing at $M_{K_s} $ = 7, and reaches its maximum
at $M_{K_s} $ = 9 or later. This implies a lack of low-luminous stars in the Hyades compared
to the field with all caution, because incompleteness in our sample starts at $M_{K_s} $ = 7.3.
In the optical regime, according to \citet{1990MNRAS.244...76K}, the maximum of the field star LF is located at
$M_V $ = 12 which corresponds to  $M_{K_s} $ = 7.3. This is at a brighter magnitude than the
Jahrei{\ss} (priv. comm.) KLF. We have no
explanation for this difference in the field star LF.

At absolute magnitudes $ 3 < M_{K_s} < 5$ we find local maxima in all  KLFs shown in Fig. \ref{lfsiggi}.
In the KLF of CNS4 we do not see a maximum but a flattening of the slope to $\alpha$ = 0.1.
In the optical this phenomenon is known as the Wielen dip \citep{1990MNRAS.244...76K}.
\citet{1990MNRAS.244...76K} located it at m = 0.6 $\rm{m}_\odot $. They explained the dip as a consequence of
an inflection of the
mass-luminosity relation which appears 
due to the formation of negative H-ions impacting the atmospheric opacity.
 A mass of 0.6 $\rm{m}_\odot $ corresponds to $M_{K_s} $ = 5.1 in the MLR  from Sect. \ref{Spatial}.
We observe strong Wielen dips (even local minima) in the luminosity functions in all regions of the Hyades
in Fig. \ref{lfsiggi}, strongest in the core region, where mass segregation is responsible
for an apparently stronger Wielen dip.

For 1 $\leq M_{K_s} \leq$ 4, we find
the slope of the luminosity function in the halo to be $\alpha$ = 0.26, i.e
marginally flatter than in young open clusters and much flatter than in the field.
This behaviour in a zone of co-moving, gravitationally unbound stars, shows that we
find there an ensemble of ``former'' members representing the IMF 
of open clusters at least in this magnitude
range. On the other hand, the slope is remarkably different from that in the field star KLF meaning that
the halo is exclusively related to the cluster and not to the field. This can be considered
as an alternative confirmation of low field star contamination (cf. Sect. \ref{contest}).

\subsection{The mass function}~\label{PDMF}
The mass function $\xi(\rm{m})$ of a cluster is defined as the number of stars
within a mass interval between m and $\rm{m} + d\rm{m}$
\begin{equation}
 \xi(\rm{m}) = \frac{dN(\rm{m})}{d\rm{m}}. 
\end{equation}
Alternatively, we can describe it as a function of the logarithm of mass

\begin{equation}
 \zeta(\log \rm{m}) = \frac{dN (\log \rm{m})}{d(\log \rm{m})} = \rm{m}\times \xi(\rm{m})\times \ln(10).
\end{equation}

\begin{figure}[h!]
   \centering
   \includegraphics[bb=62 166 582 591,angle=-90,width=7cm,clip]{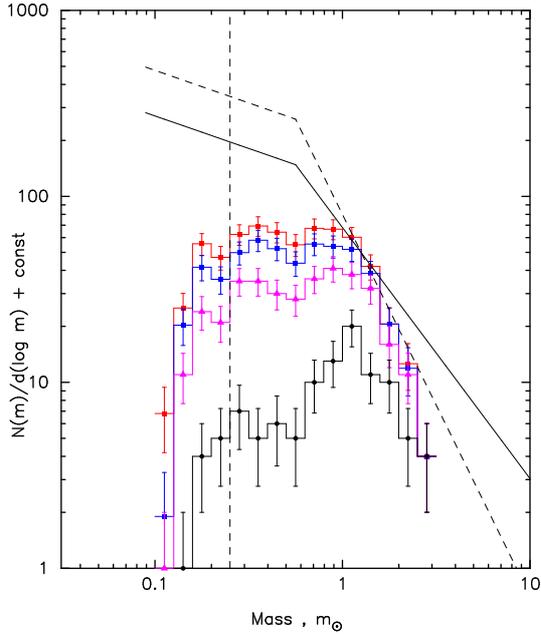}
      \caption{The present-day mass function (PDMF) of the Hyades. The lowest histogram
   (filled black dots) shows the PDMF in the core of the cluster ($\rm{r}_c < $3.1 pc),
   the magenta triangles mark the PDMF within the tidal radius ($\rm{r}_c < $9 pc),
   the blue squares stand for the PDMF in the region $\rm{r}_c < $18 pc,
   and finally the filled red squares are the PDMF of all 724 stars within 30 pc
   from the centre. The PDMFs are corrected for field star contamination. The histograms
   give the actual star counts per logarithmic mass bin of 0.1. The dashed
   vertical line at about 0.25 $\rm{m}_\odot $ marks the onset of incompleteness of our survey. For comparison,
   we show a \citet{2001MNRAS.322..231K} mass function (solid line) with $\alpha = 2.35$ for masses larger than
   0.5 $\rm{m}_\odot $, and $\alpha = 1.35$ below 0.5 $\rm{m}_\odot $.
   As an other extreme we show a mass function with slope $\alpha = 3.05$ (dashed line) fitted to
   the high-mass portion of the observed PDMF.}
         \label{mf}
   \end{figure}

\begin{figure}[h!]
   \centering
   \includegraphics[bb=62 166 582 591,angle=-90,width=7cm,clip]{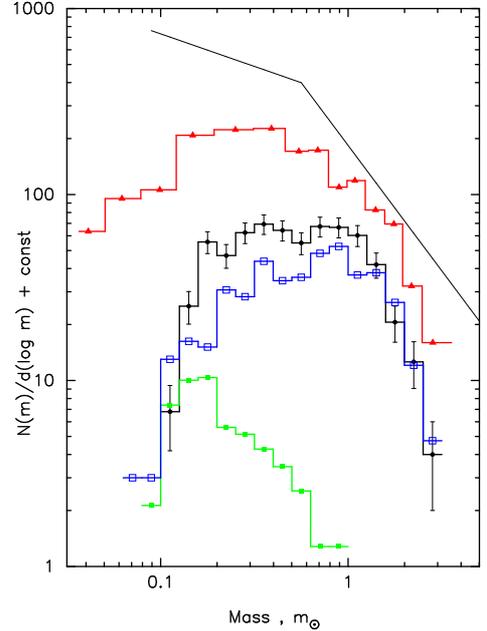}
      \caption{The mass function of the Hyades from this paper (black histogram
      with filled black dots). The blue curve (open squares) shows the
      Hyades PDMF from \citet{2008A&A...481..661B}, the red histogram with triangles
      is the PDMF from the 120 Myr old Pleiades from \citet{2003A&A...400..891M},
      and finally the green curve (filled squares) is the PDMF of Praesepe from \citet{2010A&A...510A..27B}.
      The thin solid line shows again  a \citet{2001MNRAS.322..231K} mass function.
      }
         \label{mf2}
   \end{figure}

So, $\zeta(\log \rm{m})$ gives the number of stars (systems) in the interval between
$\log \rm{m}$ and $\log \rm{m} + d\log \rm{m}$. In  Figs. \ref{mf} and
\ref{mf2} we chose  $\rm{d}\log \rm{m}$ to be 0.1.
The different regimes in the mass function $\xi(\rm{m})$ are usually approximated by a power
law with an exponent $-\alpha$ which transforms into $-\alpha+1$ for the logarithmic mass function
$\zeta(\log \rm{m})$. So, even if we show the logarithmic mass function in the figures below,
we always refer to the exponent $-\alpha$ of $\xi(\rm{m})$. 

Fig. \ref{mf} shows the mass function in different regions of the cluster. The lowest
function (black) refers to the core region ($\rm{r}_c < $3.1 pc).
The mass function peaks sharply at about 1 $\rm{m}_\odot $. Towards larger masses the MF drops
as a power law with exponent $\alpha =2.7$, even steeper than Salpeter. The most striking effect is
the steep decline ($\alpha = -0.05$) towards lower mass stars, again a clear indication for mass segregation.

The mass function for all stars within the tidal radius of the cluster $\rm{r}_c \leq  $ 9 pc has its less prominent
maximum at 0.9 $\rm{m}_\odot $, and follows about the same power law ($\alpha = 2.7$)
towards higher mass systems as in the core (provided that we neglect the four giants in the
most massive bin). Towards low-mass stars the logarithmic mass function is almost
flat ($\alpha = 0.93$) down to the completeness limit. 
This mass function refers to the gravitationally bound stars of the Hyades cluster.
Adding the formerly bound members with $\rm{r}_c \geq  $ 9 pc shifts the maximum further towards lower mass stars.
However, the logarithmic mass function for all 724 stars is almost flat 
in the mass range from 1 $\rm{m}_\odot $ to 0.2 $\rm{m}_\odot $. The mass function which we determined is incomplete
for masses lower than about 0.25 $\rm{m}_\odot $. This has, however, not much influence on the total mass of the cluster.
If we assume that the logarithmic mass function were constant below 0.25 $\rm{m}_\odot $ down to 0.01 $\rm{m}_\odot $,
this would add merely 8\% to the total mass of the cluster.

In Fig.\ref{mf} we compare the observed mass functions with
a \citet{2001MNRAS.322..231K} IMF shown as the solid line. This IMF has $\alpha = 2.35$ for masses larger than
0.5 $\rm{m}_\odot $, and $\alpha = 1.35$ below 0.5 $\rm{m}_\odot $, down to the brown dwarf limit of 0.08 $\rm{m}_\odot $.
Using the Kroupa IMF as a tangent to the observed mass function,
we can estimate a minimum initial mass of the cluster, provided that all originally formed stars
of about solar-mass are still contained within a radius of 30 pc. With these assumptions 
we get an initial cluster mass of 1100 $\rm{m}_\odot $ in 2300 stars (stellar systems). From this simple calculation
we can conclude that about 40\% of the original mass of the Hyades is still
contained  in the immediate neighbourhood of 30 pc around the centre.
The remaining mass has left further into  tails of the form predicted
by, e.g. \citet{2009A&A...495..807K}. More refined statements on the initial parameters of the Hyades
are expected from dedicated
$N$-body simulations which are underway.

A better fit to the high-mass end of the observed mass function requires a steeper slope of $\alpha = 3.05$
in a mass range from 1 to 2.5 $\rm{m}_\odot $. This is shown as the dashed line in Fig. \ref{mf}. We should
point out again that in the mass functions presented here binaries are not resolved, and the masses are
based on a specific mass-luminosity relation used to convert absolute
magnitudes to masses. This makes us cautious, not to interpret the slope of the mass function at high masses
in more depth.
%
   
In Fig. \ref{mf2} we compare our PDMF of the Hyades with earlier results. This mass function
(black histogram with filled black dots) is obtained within
the 30 pc volume around the centre (top curve in Fig. \ref{mf}).  
The PDMF from  \citet{2008A&A...481..661B} is shown as the blue histogram (open squares).
On the high mass part it is derived from the Prosser \& Stauffer (P\&S) database. \citet{2008A&A...481..661B} 
argue that the P\&S database is essentially complete down to 0.1 $\rm{m}_\odot $, but
having a 27\% contamination by field dwarfs below 0.3 $\rm{m}_\odot $. The slope of their mass function  
between 0.2 and 1 $\rm{m}_\odot $ is found to be $\alpha = 0.6$. Compared with our results this indicates that
incompleteness in the P\&S database begins near 1 $\rm{m}_\odot $ and increases towards lower masses.
We attribute this to the fact that the P\&S database is not based on a deep all-sky survey. As fainter 
stars show less concentration to the centre (mass segregation) they may be missed if the field-of-view
is restricted. 
\citet{2008A&A...481..661B} detected 3 new very low-mass stellar members, and  constructed
a PDMF down to 0.08 $\rm{m}_\odot $.
Nevertheless, their PDMF practically shows a cut-off at 0.1 $\rm{m}_\odot $. We neither
can confirm nor reject this on the basis of our analysis using the CU catalogue because of
incompleteness
below about 0.25 $\rm{m}_\odot $.
Work is going on to exploit the faint PPMXL stars not in the CU subset to reach an all-sky
completeness
of the Hyades stars at least down to 0.1 $\rm{m}_\odot $.
\citet{2008MNRAS.388..495H} did not try to derive a mass function from the 12 L-dwarfs
in their surveyed 275 deg$^2$ field. If we apply our Mass-$M_{K_s}$ relationship, the 12 candidates range
from 0.078 down to 0.06 $\rm{m}_\odot $. A larger field is needed to decide if the presently observed
strong cut-off at 0.1 $\rm{m}_\odot $ will be confirmed or not.

Fig. \ref{mf2} also shows  the PDMF of the much younger ($\approx$ 120 Myr)
Pleiades cluster from \citet{2003A&A...400..891M}. The 
maximum of the Pleiades PDMF in 
the logarithmic mass function is between 0.3 and 0.4 $\rm{m}_\odot $ compared to about 0.7 to 0.8 $\rm{m}_\odot $
in our mass function of the Hyades. If both clusters had a similar IMF then mass loss works
more effectively the
less massive the stars are. The similarity of the mass functions in the region above 1 $\rm{m}_\odot $ suggests
that this effect is less prominent for the more massive stars in a cluster. A further similarity
in the Hyades and Pleiades mass functions is the decline below 0.1~$\rm{m}_\odot $.

The green curve (filled squares) is the mass function of Praesepe
recently  derived by \citet{2010A&A...510A..27B}.
Given that Praesepe is practically of the same age as the Hyades, this mass function is quite 
unusual. With an $\alpha = 1.9$ between 0.2 and 0.7 $\rm{m}_\odot $, the contribution of low-mass stars
to the total mass of Praesepe is much more important than in the case of the Hyades and even the
Pleiades. Contrary to the latter, it would mean that Praesepe could hold its low-mass stars in bound stage.

The thin solid line in  Fig. \ref{mf2} is again the universal IMF from \citet{2001MNRAS.322..231K}. 
Prima facie one gets the impression that an IMF like Kroupa's evolves via a Pleiades PDMF
(120 Myr) to a Hyades PDMF due to dynamical and stellar evolution.

\section{The internal velocity dispersion}~\label{veldisp}
For a cluster in a {\em statistically steady state} \citep{1942psd..book.....C}
the virial theorem $ 2T = -\Omega $ holds (with T being the total kinetic and $ \Omega $
the total potential energy of the system).
For compact, isolated, massive and long living systems
like globular clusters the assumption of {\em statistically steady state} is, in
general, justified. However, this is not trivial if we were to consider a 
relatively old, low-mass open cluster dipped into the Galactic disk.  

Strictly speaking, the virial theorem only holds for the cluster as an ensemble. For each
well-defined subset of stars $\{m_1, .....,m_s\}$ we can postulate that the velocity dispersion
is given by
\begin{equation}
<\rm{v}^2> = 0.5 G \sum \frac{m_i m_j}{r_{i,j}} \left /  \sum m_i \right .
\label{graven}
\end{equation}
where $m_i$ is the mass of an individual star in the subset,
and $m_j$ runs along all stars within the tidal radius of the
cluster,  $r_{i,j}$ is the distance between stars i and j,
G is the gravitational constant,
and $<\rm{v}^2>$ is defined as
\begin{equation}
<\rm{v}^2> \times \sum m_i = \sum m_i \rm{v_i}^2
\end{equation} 
The
summation in Eq. \ref{graven} extends over all pairs of stars with i $\neq$ j.
If the subset consists of all stars within the tidal radius then Eq. \ref{graven}
gives $ 2T = -\Omega $. 

We use here the actual three-dimensional distribution of the Hyades stars and do not 
approximate the cluster by a Plummer model, albeit the fit of the Hyades to
a Plummer model is excellent (see Sect.\ref{Spatial}).
The assumption of a {\em statistically steady state},
if at all, is only justified within the tidal radius of the cluster. We applied
Eq. \ref{graven} to two distinct subsets of the Hyades, one containing all stars within
the core radius $\rm{r_{co}}$ ($\rm{r_c} \le 3.1$~pc) and the other within the 
tidal radius $\rm{r_t}$ ($\rm{r_c} \le 9.0$~pc).
The three-dimensional velocity dispersion comes out as 0.42~${\rm km s^{-1}}$ and 
0.32 ${\rm km s^{-1}}$, respectively.
Putting an artificial particle in the centre we obtain the central velocity dispersion
$<\rm{v}^2>^{1/2}$ = 0.45 ${\rm km s^{-1}}$.

\citet{1988AJ.....96..198G} used their fit of the Hyades cluster to a Plummer 
model (cf. Sect.\ref{Spatial}) 
to get 0.51  $\rm{km s^{-1}}$ for the central, and 0.40 $\rm{km s^{-1}}$ for the average
velocity dispersion (or 0.23 $\rm{km s^{-1}}$ for 1-D). The higher values can be 
explained by their assumption of a higher (tidal) mass for the Hyades.
\citet{1998A&A...331...81P} derive a central
velocity dispersion of 0.36 ${\rm km s^{-1}}$ from their fit to a Plummer model, which is too low
because of their lower mass estimate for the cluster.

In principle, an empirical determination of the velocity dispersion is
possible by exploiting the observed velocity distribution of cluster members. However,
during their life, open clusters are permanently losing members part of which remain in the vicinity of their
parent cluster and form sparse haloes and tails of co-moving stars. This makes it difficult to
separate ``actual'' members by use of velocity observations and properly determine $<\rm{v}^2>^{1/2}$. 
For open clusters, one expects velocity dispersions ranging
from a few hundred meters per second to 1-2~${\rm km s^{-1}}$  i.e., at a level of  accuracy
of radial velocity measurements and/or tangential velocities of nearby stars. Empirically determined $<\rm{v}^2>$ are thus
sensitive to the adopted mean errors of input data used for computing internal
spatial velocities of cluster members \citep[see e.g., Appendix in][]{2007A&A...468..151P}.

\begin{figure}[h!]
   \centering
   \includegraphics[bb=93 53 332 635,angle=-90,width=8cm,clip]{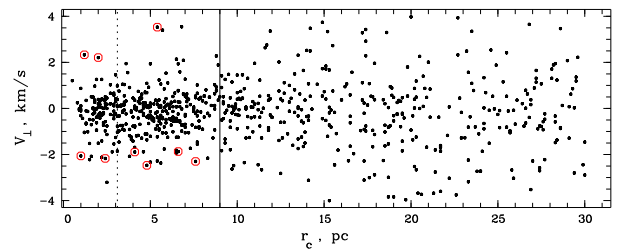}
      \caption{The distribution of the tangential velocity components  $\rm{v}_\perp$ perpendicular to the direction to the 
convergent point as a function of the distance from the cluster centre. The nine stars in red circles are outliers which are not
considered in the determination of the velocity dispersion in Table \ref{table:3}.
The vertical lines mark the core (dashed) and tidal (solid) radii.}
         \label{vd}
   \end{figure}

In the past, several attempts were undertaken to derive the
velocity dispersion of the Hyades from proper motions and radial velocities.
\citet{1990A&A...228...69S} determined the velocity dispersion from 
44 FK5 stars. Using $\rm{v}_\perp$, he obtained an 1d-velocity dispersion of
0.71 $\pm$ 0.41~${\rm km s^{-1}}$. Based on only the $\rm{v}_\alpha$-component,
the velocity dispersion comes out to be 0.3~${\rm km s^{-1}}$ which corresponds
to about 0.5~${\rm km s^{-1}}$ for the three-dimensional dispersion. \citet{1990A&A...228...69S}
explained this considerable difference by larger uncertainties in 
$\rm{v}_\delta$-component of the velocity data. \citet{1998A&A...331...81P} used the Hipparcos parallaxes
and proper motions, but were unable to determine the velocity dispersion,
because the accuracy of the measured trigonometric parallaxes resulted in a too large mean error
of the tangential velocities. \citet{2001A&A...367..111D} used the same data set as \citet{1998A&A...331...81P}
but took the formally better secular parallaxes instead of the trigonometric parallaxes.
They achieved  a 1d-velocity
dispersion of 0.3~${\rm km s^{-1}}$ after consecutive rejection of stars with large offsets
in the observed vs. computed tangential velocities.
In the following we estimate the three-dimensional velocity dispersion
in the Hyades via the one-dimensional velocity dispersion given by the velocity component perpendicular to the direction to the 
convergent point $\rm{v}_\perp$. The distribution of $\rm{v}_\perp$ versus the distance
of stars from the cluster centre is shown in Fig. \ref{vd}. Stars in red circles are outliers and are discussed
a few paragraphs below.

Let $\rm{v}_{\it{i}}$  be the residual space velocity of the $i$-th member
with respect to the bulk motion of the cluster,
and $\sigma^2_{\it{i}}$  its individual mean error. We define the square of the velocity dispersion
${<\rm{v}^2>}$ as
\begin{equation}
<\rm{v}^2> = \frac{1}{n}\sum\rm{v}^2_{\it{i}} - \frac{1}{n}\sum^{}\sigma^2_{\it{i}}
   \label{v3d}
\end{equation}
where $n$ is the number of stars in a given region.
Provided that the space velocities of the cluster stars are randomly oriented, 
the first term on the right-hand side can be computed from the velocity components $\rm{v}_\perp$
and we can re-write Eq. \ref{v3d} as
\begin{equation}
V^2_d = <\rm{v}^2> = 3 \times (<\rm{v}_\perp^2> - <\sigma_\perp^2>).
   \label{v1d}
\end{equation}
The mean errors $\sigma_{V_d}$ \citep{1953stas.book.....T} are computed as
\begin{equation}
\sigma_{V_d} = \frac{3\sum\rm{v}^2_{\it{i}}}{V_d(2n)^{1/2}}.
\end{equation}
Here we are obliged to rely on the quoted $\sigma_{\it{i}}$
and to assume their uncertainties to be negligible.

Since the kinematic candidates of the Hyades have different observing history, their
individual mean errors $\sigma_\perp$ vary significantly, from $\approx 0.1\,\rm{km s^{-1}}$
for bright  (and  high mass) Hipparcos stars to $\approx 1\,{\rm km s^{-1}}$ for the
faintest (and lowest mass) stars in our sample. Therefore, they must be considered in different
bins containing a sufficient number of stars of homogeneous accuracy. As a compromise,
we divided our sample into four groups, according to the masses of stars included:
masses higher than 1.45~$\rm{m}_{\odot}$, between 1.45~$\rm{m}_{\odot}$ and 
1.05~$\rm{m}_{\odot}$, between 1.05~$\rm{m}_{\odot}$ and 0.71~$\rm{m}_{\odot}$,
and less than 0.71~$\rm{m}_{\odot}$. We  had to exclude the last group from 
consideration: there is no chance to use them for the determination of the velocity 
dispersion due to the low accuracy of their kinematic data. Further we rejected 
5 stars from  the calculation of
$<\sigma_\perp^2>$ because their reported mean errors were significantly
higher than the corresponding group average.

For each mass group, the mean values $<\rm{v}_\perp^2>$ were computed in two ranges of
the distance from the cluster centre: $\rm{r_c} \leq 3.1$~pc, and 
$\rm{r_c} \leq 9$~pc.
It is self-evident that stars with high $\mid\rm{v}_\perp\mid$
can cause an oversize increasing of the velocity dispersion. Therefore, in each bin
we applied a statistical test to reveal outliers at the 95\% confidence level.
In total, 9 stars were excluded from the calculation of the velocity dispersions. They
are marked by red circles in Fig. \ref{vd}. All these stars are double stars. 
Of these, six are found as $\Delta\mu$-binaries in \citet{1999A&A...346..675W}.
The binary nature of the $\Delta\mu$-stars is revealed from the significant difference
between their short-scale (Hipparcos) and their long-scale (FK5, Tycho-2)
proper motions. One more  $\Delta\mu$-binary is found in \citet{2007A&A...464..377F}.
Another star is marked as double in the Hipparcos Double and Multiple Systems Annex,
and the last one is WDS 04285+1742 from the 2011 version of the Washington Double Star catalog
\citep{2001AJ....122.3466M}.

For comparison, we also computed the velocity dispersion of stars at larger distances 
from the cluster centre by applying the same approach. Since there are only 8 stars 
at $\rm{r_c}>$9~pc with masses larger than 1.45~$\rm{m}_{\odot}$, the calculations were carried 
out only for two mass groups.
The results are given in Table \ref{table:3} together with the theoretical velocity dispersion
computed above from the virial theorem.
\begin{table}
\caption{Three-dimensional velocity dispersions derived from the potential energy (Eq. \ref{graven}) as well as computed
from the observed velocities of stars in 3 mass ranges: I ($\rm{m} \geq 1.45\,\rm{m}_{\odot}$), II
($1.05\,\rm{m}_{\odot} \leq \rm{m} < 1.45\,\rm{m}_{\odot}$), III ($0.71\,\rm{m}_{\odot} \leq \rm{m} < 1.05\,\rm{m}_{\odot}$).}             
\label{table:3}      
\centering                          
\begin{tabular}{c c c c c }        
\hline\hline                 
                & theoretical & I & II &  III \\
$\rm{r_c}$  & $<\rm{v}>^{1/2}$ & $<\rm{v}>^{1/2}$ & $<\rm{v}>^{1/2}$ & $<\rm{v}>^{1/2}$ \\
pc        &$\rm{km s^{-1}}$ & $\rm{km s^{-1}}$ & $\rm{km s^{-1}}$ & $\rm{km s^{-1}}$ \\          
\hline 
  0        & 0.45 &                 &                 &                 \\
$\leq 3.1$ & 0.42 & 0.81 $\pm$ 0.13 & 0.76 $\pm$ 0.15 & 0.74 $\pm$ 0.21 \\
$\leq 9.0$ & 0.36 & 0.77 $\pm$ 0.10 & 0.88 $\pm$ 0.11 & 0.82 $\pm$ 0.13 \\
\hline
$ > 9.0$  &       &     & 2.68 $\pm$ 0.38 & 2.38 $\pm$ 0.24 \\
\hline
\end{tabular}
\label{vdisp}
\end{table}

The increase of the velocity
dispersion outside 9 - 11~pc from  the cluster centre can be already observed in Fig. \ref{vd}, and
Table~\ref{table:3} confirms this impression. Partly, this may be explained 
by an increasing contamination by field stars.  
But more probably, this finding indicates the limiting 
distance for the gravitationally bound part of the Hyades cluster. The distance 9 - 11~pc
is in good agreement with the conclusion on the cluster size
we obtained from the density and mass distribution of stars in Sec. \ref{Spatial}. 

Within the first two bins ($r_c \leq 9$~pc), the velocity dispersion achieves 
$\approx 0.8 \,\rm{km s^{-1}}$ which is higher by almost a factor of two than one would
expect from Eq. \ref{graven}. So, a factor of 4 in mass would be required to balance this
discrepancy.
Assuming additional mass
hidden in double stars, white dwarfs, faint stars below the completeness limit of our survey,
the current estimate of cluster mass of approximately $ 300\,\rm{m}_{\odot}$ can be hardly
increased actually by more than 50\%, a factor of 4 is excluded.   

On the other hand, we can assume that $<\rm{v}_\perp^2>$ 
is overestimated and/or  $<\sigma_\perp^2>$ is
underestimated in Eq. \ref{v1d}. A formal calculation shows that a decrease of the
observed velocity dispersion by a factor of two can be achieved for the most massive stars (group I)
by increasing the rms errors of
proper motions by a factor of $\approx 3$. However, this seems to be far from reality, especially 
for this mass group containing mainly stars
with Hipparcos data.

Alternatively, an additional dispersion can be introduced by the presence of double stars
which can impact the proper motion measurements. The effect from neglecting the
binary nature of the stars is twofold: the velocity dispersion contains a part that comes
from orbital motion, and the total gravitational energy is increased to the higher mass
in binaries. Indeed, we found that the high velocity
dispersion is caused by only 20\% of the stars with somewhat higher $\rm{v}_\perp$ 
velocities. From a statistical point-of-view, however, they 
cannot be considered as outliers and simply be excluded from the
calculation of the velocity dispersion. But, we note that this number of 20\% coincides well with 
the results by \citet{1988AJ.....96..198G}, who estimated the percentage of double stars to be 25\% from
their measurements
of radial velocities of stars in the Hyades region and extrapolated this number
to the complete cluster. As a further possibility to explain the high empirical
velocity dispersion we can assume that a certain number of stars 
is just escaping the cluster.
Simulations have been started to quantitatively investigate the importance of this effect.

In the discussion above we used the observed distribution of stars (with their masses) 
to derive the allowed velocity dispersion for the cluster to be in equilibrium. That is,
the mass of the cluster was obtained by counting the members with their individual masses.
At first sight, see Table \ref{vdisp} the observed velocity dispersion seemed to considerably
exceed the allowed one. But, as we argued above, there are good reasons that both the observed
distribution of masses and the observed velocity dispersion are consistent with each other.

Let us, at the end of this chapter, add a remark on the estimation of masses from an observed
velocity distribution. \citet{1942psd..book.....C} introduces 
the ''average'' radius R of the cluster such that the equation
\begin{equation}
M <\rm{v}^2> = \frac{G}{2} \frac{M^2}{R}
\end{equation}
holds. Here M is the total (gravitationally bound) mass of the cluster. This is fulfilled for 
\begin{equation}
\frac{1}{\rm{R}} =\frac{1}{M^2}\sum \frac{m_i m_j}{r_{i,j}}
\end{equation}
where the summation is now carried out over all $m_i, m_j$ with $i\neq j$. The so-called ''average''
radius of the cluster is determined via the weighted mean over the inverse distances between
individual stars. For the Hyades with their tidal mass of 276 $m_{\odot}$ we find R = 4.53 pc,
which in our case happens to be 1.5 times $r_{co}$. The measured velocity dispersion can be
used to determine the mass (virial mass) of an open cluster via
\begin{equation}
M  = \frac{2}{G} <\rm{v}^2> R.
\label{virialmass}
\end{equation}
The two terms on the right-hand side of Eq.\ref{virialmass}, $<\rm{v}^2>$ and R depend on the sample for
which the velocity dispersion is measured.
This should be taken into account when deriving the mass of a cluster from its velocity dispersion.

\section{Summary}~\label{summary}
We used a subset of PPMXL, derived from a combination
of PPMXL proper motions with UCAC3 and CMC14 observations,
and augmented by CMC14 photometry to perform the most complete
three dimensional census of the Hyades to date. This census yielded 724 stellar systems
as candidates moving with the same space motion as the cluster.
These candidates have been found from an application of the convergent
point method, and we verified the resulting secular parallaxes
by photometric parallaxes. Although we
tested the full catalogue for stars co-moving with the Hyades centre, we restricted
our analysis to stars with derived distances smaller than 30 pc from the cluster centre.
We did not search for new white dwarfs, as they may be at the detection limit of the 2MASS,
and need a separate approach.
Our census is complete down to $M_{K_s} \approx$ 7.3, corresponding to masses of 0.25  $\rm{m}_\odot $.
Incompleteness for more massive stars is no longer a question of limited field-size; the only reason
for incompleteness may be possibly erroneous proper motions and/or photometry.

The secular parallaxes allow us to construct an empirical Colour-Magnitude Diagram (here $ M_{K_s}$
vs. $J-K_s$) from the brightest stars down to $M_{K_s} \approx$ 9.  
Comparing this with 3 different theoretical isochrones relevant to the metallicity and
age of the Hyades we find that only the isochrones from \citet{1998A&A...337..403B} more or less
describe the loci of the Hyades stars fainter than about $M_{K_s} =$ 5 in the CMD, whereas the isochrones 
from \citet{2008A&A...482..883M} and \citet{2008ApJS..178...89D} predict  $J-K_s$ colours which are too blue.

Using the mass-luminosity relation described in Sect. \ref{Spatial} we derive M(r), the cumulative mass 
as a function of the distance from the cluster centre. M(r) is increasing up to our arbitrarily set
limiting radius of 30 pc. If we consider the tidal field of the Galaxy via Eq. \ref{tidal},
we derive a tidal radius of about 9 pc, consistent with earlier determinations.
The M/L relation gives us system masses for unresolved binaries. Using these individual
masses we find that 276 $\rm{m}_\odot $ (364 stars) are bound within the tidal radius of the cluster,
another 100 $\rm{m}_\odot $ (190 stars) are found in the halo up to 2 tidal radii,
and another 60 $\rm{m}_\odot $ (170 stars)
are found as co-movers outside 2 tidal radii. Here the masses are corrected for contamination,
whereas the number of stars is given as counted. 

The spatial distribution of the Hyades has a clear ellipsoidal shape considering both bound as
well as unbound stars.  
The largest axis is almost parallel to the galactic plane, and
forms an angle of about 30 deg with the galactocentric radius vector (in the direction
of the galactic rotation). The ellipticity and the orientation of the principal axis may already reflect
the onset of tidal tails as predicted by the models.

Within its tidal radius  ($\rm{r_t = 9 pc}$) the spatial distribution follows excellently
a Plummer model with core radius $\rm{r_{co} = 3.1 pc}$ and a central density of 2.21 $\rm{m_{\odot} pc^{-3}}$.
The half-mass radius of the cluster is $\rm{r_h = 4.1 pc}$, and the ratio $\rm{r_h}/\rm{r_{co}}$
is 1.32, also in excellent agreement with the model value. The ``average'' cluster radius R determined
from the distances between the stars is found to be 4.53~pc. 

Mass segregation is clearly seen in the Hyades. The average mass per star drops from 1.1 $\rm{m}_\odot $
at the centre to 0.5 $\rm{m}_\odot $ at the tidal radius.
This mass segregation is also seen in the $K_s$-band luminosity function KLF.
In the core the KLF shows  a peak at $M_{K_s} \approx $ 4 which is more prominent  
than the other at $M_{K_s} = $ 7, whereas this is vice versa in the corona. We see clear indications
for the presence of the so-called Wielen dip at about $M_{K_s} = $ 5. Mass segregation
exaggerates the Wielen dip in the core, while it is present in all other regimes (corona, halo
and co-movers) as a flattening of the slope of the luminosity function.

Compared with the KLF of the field stars in the solar neighbourhood (Fig. \ref{Fig3})
the Hyades KLF is generally flatter than the field KLF in all regions of the cluster
in the range  1 $ < M_{K_s} < $ 6. The increase in the field KLF for $M_{K_s} > $ 7 is
not seen in the Hyades KLF with the caution that our sample is becoming incomplete there.

The mass function of all co-moving stars (bound or unbound) is complete down
to 0.25 $\rm{m}_\odot $. About 94\% of the total mass is in stars
more massive than 0.25 $\rm{m}_\odot $. Hypothetically, even a constant logarithmic mass function
at lower masses between 0.25 $\rm{m}_\odot $ and 0.01 $\rm{m}_\odot $
(for which we so far see no indication) would add only
8\% to the total mass we determined. From literature there is no indication for a considerable amount
of mass in white dwarfs. Hence, hidden mass could only be in unresolved binaries, and, indeed,
we got hints to this from the observed high velocity dispersion. 
Adopting the estimate of 25\% binaries from \citet{1988AJ.....96..198G}, the total mass would
increase by
17\%.

Mass segregation is also seen impressively in the mass function (Fig. \ref{mf}). Whereas in the core,
the PDMF sharply peaks at about 1 $\rm{m}_\odot $ and steeply drops to both sides, the number of
lower mass stars increases up to the tidal radius to give a practically flat logarithmic
mass function. If the IMF had been a \citet{2001MNRAS.322..231K} IMF we see strong dynamical
evolution on the low-mass side. This would hold even more for a Salpeter IMF.
A comparison with the mass function of a much younger cluster like the Pleiades supports this
conclusion. Adopting a Kroupa IMF we derive a minimum mass of 1100 $\rm{m}_\odot $ for the cluster
at formation.

The distribution of the velocity dispersion is clearly different inside and outside the tidal
radius of the cluster.
The observed velocity dispersion within the tidal radius is not sustained by the total
gravitational energy of the cluster. A relatively high fraction of binaries is needed
to overcome this discrepancy, because binaries not only increase the total mass, but their
orbital motions also may strongly corrupt the proper motions used for the determination
of the velocity dispersion.
Our investigation on possible contamination by local field stars has
shown that the population outside the tidal radius is strongly related to the
cluster itself and consists of stars that left the cluster in the past.

Putting all together, we observe our neighbouring open cluster as an
evolving object, which is permanently losing bound members that we can pin-point
up to a certain distance once they left the cluster. The results that we obtained here
for the spatial, the velocity and the mass distribution of the Hyades give starting
conditions for ongoing $N$-body simulations to reveal the past and the future of this cluster 

\begin{acknowledgements}
Part of this work was supported by DFG grant RO 528/10-1.
We are grateful to Hartmut Jahrei{\ss} who furnished us with the KLF of the nearby stars
prior to publication. We also thank Hans Zinnecker and Andreas Just for helpful
discussions. We further thank the anonymous referee for his helpful comments.  
This paper is based on
observations from the ESA Hipparcos satellite.
This publication makes use of data products from the Two Micron All Sky Survey,
which is a joint project of the University of Massachusetts and the
Infrared Processing and Analysis Center/California Institute of Technology,
funded by the National Aeronautics and Space Administration and the National Science Foundation.
This research has made use of the SIMBAD database,
operated at CDS, Strasbourg, France.
\end{acknowledgements}

\end{document}